\documentclass[aps,prl,twocolumn,superscriptaddress,longbibliography]{revtex4-1}%
\usepackage{graphicx} 
\usepackage{float} 
\usepackage{dcolumn}
\usepackage{bm,color}
\usepackage{amsmath,amssymb,dsfont,amstext,amsfonts}
\usepackage{extarrows}
\usepackage[colorlinks=true,linkcolor=blue,urlcolor=blue,citecolor=blue]{hyperref}
\usepackage{xcolor}
\usepackage{wasysym}
\usepackage{mathtools}
\usepackage{bbold}
\usepackage[normalem]{ulem} 
\usepackage{color}

\newcommand{\bk}{\bm{k}}

\usepackage{setspace}

\usepackage{titlesec}
\titleformat*{\section}{\normalsize\bfseries\centering}

\makeatletter
\def\l@subsection#1#2{}
\def\l@subsubsection#1#2{}
\makeatother
\usepackage{inputenc}

\begin{document}
\title{Floquet multi-gap topology: Non-Abelian braiding and anomalous Dirac string phase}
\author{Robert-Jan Slager}
\affiliation{TCM Group, Cavendish Laboratory, University of Cambridge, JJ Thomson Avenue, Cambridge CB3 0HE, United Kingdom\looseness=-1}
\author{Adrien Bouhon}
\affiliation{TCM Group, Cavendish Laboratory, University of Cambridge, JJ Thomson Avenue, Cambridge CB3 0HE, United Kingdom\looseness=-1}
\author{F. Nur \"{U}nal}
\affiliation{TCM Group, Cavendish Laboratory, University of Cambridge, JJ Thomson Avenue, Cambridge CB3 0HE, United Kingdom\looseness=-1}
\date{\today}

\begin{abstract}
Topological phases of matter span a wide area of research shaping fundamental pursuits 
and offering promise for future applications. While a significant fraction of topological materials has been characterized using symmetry requirements of wave functions, the past two years have witnessed the rise of novel multi-gap dependent topological states, the properties of which go beyond these approaches and are yet to be fully explored. Thriving upon these insights, we report on uncharted anomalous phases and properties that can only arise in out-of-equilibrium Floquet settings. In particular, we identify 
Floquet-induced non-Abelian braiding mechanisms, which in turn lead to a phase characterized by an anomalous Euler class, the prime example of a multi-gap topological invariant. Most strikingly, we also retrieve the first example of an `anomalous Dirac string phase'.
This gapped out-of-equilibrium phase features an unconventional Dirac string configuration that physically manifests itself via anomalous edge states on the boundary. Our results therefore not only provide a stepping stone for the exploration of intrinsically dynamical and experimentally viable multi-gap topological phases, but also demonstrate a powerful way to observe these non-Abelian processes notably in quantum simulators.
\end{abstract}
\maketitle

Topological band insulators and semimetals profit from connections to mathematical principles to characterize new physical behaviors in tangible systems~\cite{Rmp1,Rmp2,Weylrmp}. Although a plethora of results have been established~\cite{clas1,clas2,Vanderbilt_smooth_gauge, Cornfeld_2021, InvTIVish, Shiozaki14, Ahn2019, Alex_BerryPhase}, these by and large depend on symmetry representations of wave functions over the Brillouin zone (BZ) torus and have been captured in systematic schemes that can address a significant part of such symmetry protected crystalline free fermion phases~\cite{clas3,clas4,clas5}. In the past two years, however, a set of more exotic phases that depend on multi-gap conditions~\cite{bouhon2020geometric} and, hence, cannot be addressed by previous characterizations, has been gaining interest~\cite{Wu1273, bouhon2019nonabelian, BJY_nielsen}. Rather, they arise by braiding band degeneracies~\cite{Wu1273, bouhon2019nonabelian, BJY_nielsen} that, similar to disclinations in bi-axial nematic phases \cite{Kamienrmp, volovik2018investigation, Beekman20171}, can carry non-Abelian frame charges. Braiding such {\it singular points} in momentum space renders similarly-valued charged degeneracies between two bands possible, which in turn is quantified by a topological number known as Euler class, signifying the obstruction to their annihilation. Theoretical predictions~\cite{Peng2021,peng2022multi, park2022nodal, Park2021, Lange2022,chen2021manipulation, magnetic, Wieder_axion} have further culminated in material predictions~\cite{Peng2021,peng2022multi, bouhon2019nonabelian} as well as recent observations in meta-materials~\cite{jiang_meron,Guo1Dexp,Jiang2021,qiu2022minimal}, while the prediction of topologically stable monopole--antimonopole pairs imprinted upon quenching by a non-trivial Euler Hamiltonian~\cite{Unal_quenched_Euler} has been promptly followed by experimental verification~\cite{zhao2022observation}. These rapid developments as well as their fundamentally unique nature have promoted multi-gap topological pursuits as a promising research direction involving various platforms that range from phonons~\cite{Peng2021, Park2021} and synthetic systems to strained electronic systems~\cite{chen2021manipulation, magnetic}.

Going beyond static phenomena, study of topology in out-of-equilibrium settings has not only 
revealed new topological classification schemes and new connections between different invariants~\cite{Roy17_PRB,Kitagawa10_PRB,Rudner13_PRX,Unal19_PRR,Unal16_PRA,Wangchern_17_PRL,GoldmanDalibard14_PRX,Eckardt17_RMP,Cooper19_RMP,HuZhao20_PRL}, but also offered promising handles as dynamical probes in optical lattices~\cite{Wangchern_17_PRL,WangUnal_18_PRL,RaciunasUnal_18_PRA}.
In this regard, Floquet systems stand out with the periodic nature of their spectrum, where the {\it quasienergy} can be defined solely up to modulo $2\pi$, forming the Floquet Brilloun zone (FBZ) which repeats itself in integer multiples of $2\pi$~\cite{GoldmanDalibard14_PRX,Eckardt17_RMP,Cooper19_RMP}.
Essentially, one obtains quasienergy bands with a crucial difference, compared to static counterparts, of one additional gap at the FBZ edge connecting the replicas of bands. With the possibility of harbouring edge states also in this extra gap~\cite{Kitagawa10_PRB,Rudner13_PRX}, anomalous Floquet topological insulators have attracted great attention, where the Chern number has been rendered insufficient to predict edge states, triggering also experimental pursuits and verifications~\cite{Wintersperger20_NatPhys,Unal19_PRL,Tarnowski19_NatCom,Maczewsky17_NatCommun,Mukherjee17_NatComm}.

\begin{figure}
	\centering\includegraphics[width=.95\linewidth]{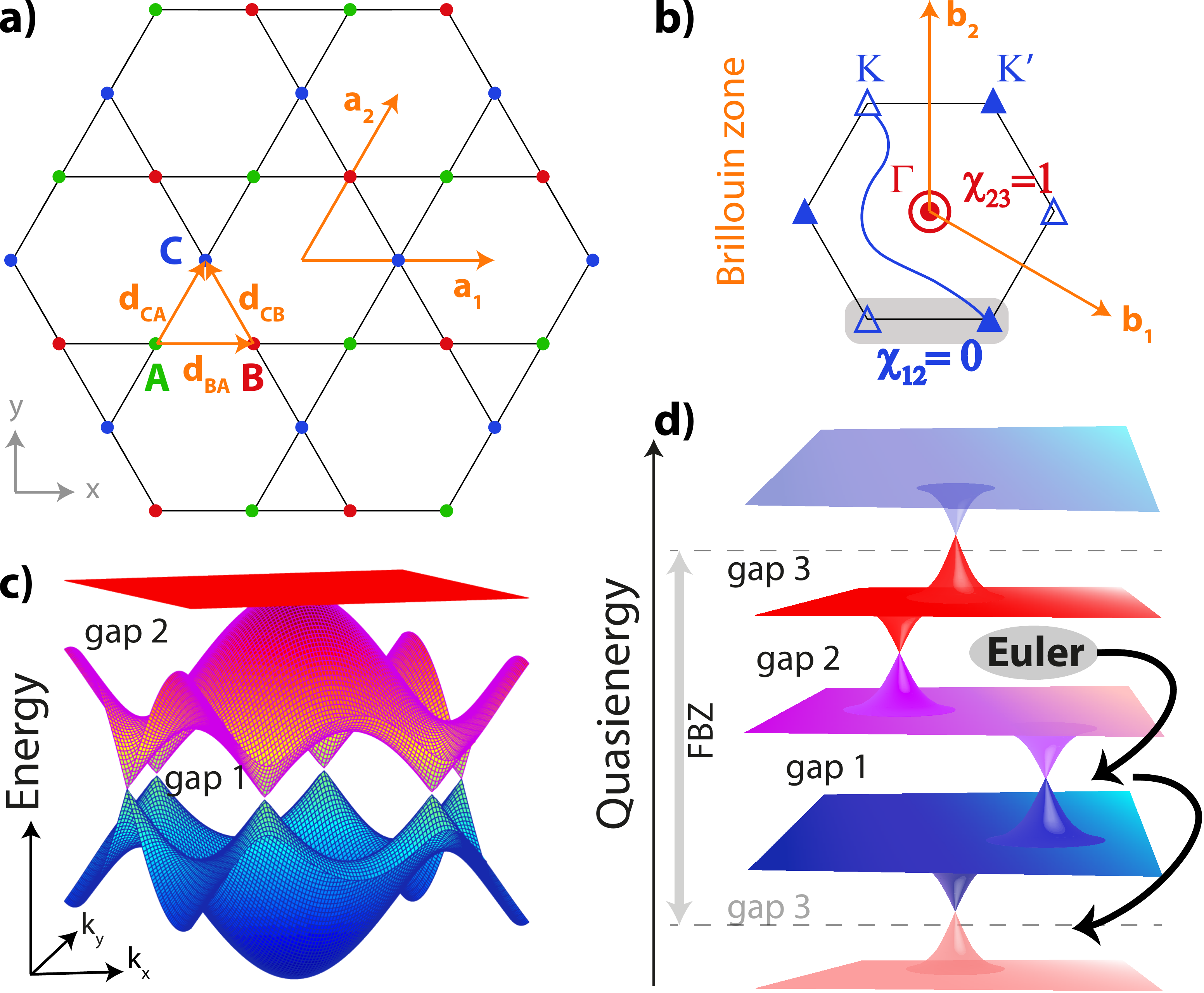}
	\caption{{\bf Static Kagome model and multi-gap topology in Floquet spectrum. } a) The Kagome lattice underlying model (\ref{eq:Hamiltonian}). b) BZ harbours multi-gap topological configuration of band nodes visible in the static band structure (c) for vanishing $\Delta_{\alpha}$. The double node between bands 2 and 3 (red circles) at the $\Gamma$-point is stable and characterized by patch Euler class $\chi_{23}=1$, while the nodes at $K$ and $K'$ (triangles) in gap 1 are connected by a Dirac string (blue line). These nodes can be annihilated (illustrated with empty/filled markers), hence, have patch Euler class (shaded)  $\chi_{12}=0$. d) A schematic Floquet spectrum. The periodic nature of quasienergy culminates in an extra gap (3) at the egde of the Floquet Brillouin zone. As a result, non-Abelian braiding of band nodes can involve any of these band gaps.
	}
	\label{fig1}
\end{figure}

Here we address the fundamental question whether multi-gap topological phases beyond equilibrium counterparts exist in Floquet settings and investigate the role of anomalous gap on the Euler topology as well as relevant physical signatures. We find that these questions can be answered affirmatively. 
In particular, band nodes can be braided between bands [which we will denote as `gaps' despite being gapless due to the presence of the degeneracies] within the FBZ as well as over the edge of FBZ as illustrated in ~Fig.\ref{fig1}d. This similarly holds for Dirac strings -- lines marking a change in gauge (from plus to minus sign) that connect band nodes in a gap to indicate flipping of charge as will be detailed in the subsequent-- that can reside in between the bands of different FBZs. As a result, we present new braiding processes involving all gaps in the quasienergy spectrum, which are by definition anomalous and can only exist in periodically driven systems. Moreover,
we demonstrate a fully gapped phase with all nodes removed via braiding, while the system still hosts anomalous Dirac strings allowed by the Floquet spectrum, which 
physically correspond to unaccounted edge states, providing for a direct observable to distinguish this new anomalous phase.

{\it Multi-gap Floquet topology--}
When a system has ${\cal C}_2{\cal T}$ [combination of two-fold rotations and time reversal] or ${\cal PT}$ [parity and time reversal] symmetry, band nodes carry non-Abelian frame charges with respect to those in the adjacent gaps~\cite{bouhon2019nonabelian} due to the admitted real representation of the Hamiltonian~\cite{Wu1273,BJY_nielsen, bouhon2019nonabelian}, acting as the analogue of disclinations in bi-axial nematic phases \cite{Kamienrmp, volovik2018investigation, Beekman20171}. Braiding such {\it singular points} in momentum space renders similarly-valued charged degeneracies between two bands possible. A particular insightful perspective to quantify the emergent multi-gap topology is attained by considering Dirac strings that can efficiently trace non-Abelian topological phase transitions~\cite{Jiang2021,Peng2021}, which will be useful in our characterization of Floquet systems below. Each of the nodes within a certain gap is connected by a Dirac string (blue line in Fig.~\ref{fig1}b) and  when a node crosses the Dirac string of a pair of nodes in a neighboring gap, its frame charge changes sign. As a result, the momentum space braiding of nodes residing in adjacent gaps may culminate in a phase where, within a certain gap (i.e., in a band subspace), band degeneracies have same-valued frame charges. The corresponding topological obstruction is then quantified by the Euler class evaluated over a BZ patch that only contains the relevant nodes via the non-Abelian Berry curvature in their band subspace (see Ref.~\cite{bouhon2019nonabelian} and Methods).

With the important role played by adjacent gaps in multi-gap topology, it is compelling to address the effect of a Floquet drive and the periodicity of the spectrum~\cite{GoldmanDalibard14_PRX,Eckardt17_RMP}. Under a periodic modulation with frequency $\omega=2\pi/T$, the Floquet spectrum is resolved as the phase eigenvalues ($e^{-i\varepsilon T}$) of the time evolution operator evaluated over a full period, $U(T)=\mathfrak{T}\exp\{-i\int_0^T H(t)dt\}$, for time ordering $\mathfrak{T}$ and the Hamiltonian $H(t+T)=H(t)$. The quasinergy $\varepsilon$ is defined on a circle, where we identify the FBZ with $\varepsilon T\in(-\pi,\pi]$ as marked in Fig.~\ref{fig1}d. In the context of Euler class, we here consider periodic drives that preserve the ${\cal C}_2{\cal T}$ symmetry to ensure that the Floquet eigenstates (solutions of the time-dependent Schr\"odinger equation) can be brought into real form together with the non-Abelian frame charges of the nodes~\cite{Wu1273,BJY_nielsen,bouhon2019nonabelian}. 
Crucially, this renders popular circular drives inapt in common ${\cal C}_2$-symmetric setups, since the chirality breaks $\cal{T}$and we naturally assume symmorphic time reversal symmetry~\cite{magnetic}. As a result, two main routes emerge that either involve using linearly-polarized driving or a circular drive that is rotating one direction for half a cycle and in the opposite direction for the other half.

{\it Model setting--} To illustrate our findings we depart from a simple Kagome geometry, given in Fig.~\ref{fig1}, which we stress merely serves as model setting with ${\cal C}_2{\cal T}$ and in no matter affects the generality of the results presented. Taking into account nearest neighbor (nn) hopping terms with amplitude $J$ as well as on-site potentials $\Delta$, the Hamiltonian, diagonal in crystal momentum $\bk$, is written as
\begin{equation}\label{eq:Hamiltonian}
H(\bk)= -2J \sum_{\alpha\neq\beta}\cos(\bk \bm{d}_{\alpha\beta}) c^{\dagger}_{\alpha}c_{\beta} +\sum_{\alpha}\Delta_{\alpha}c^{\dagger}_{\alpha}c_{\alpha} ,
\end{equation}
where $c^{(\dagger)}_{\alpha}$ is the annihilation (creation) operator at site $\alpha$ for $(\alpha,\beta)\in(A,B,C)$ denoting the three orbital basis, with the nn distances given by $\bm{d}_{BA}=\frac{1}{2}\hat{x}, \, \bm{d}_{CA}=\frac{1}{4}\hat{x}+\frac{\sqrt3}{4}\hat{y}$ and $\bm{d}_{CB}=-\frac{1}{4}\hat{x}+\frac{\sqrt3}{4}\hat{y}$ [see Fig.~\ref{fig1} and Supplemental Material (SM)]. Hereafter, we set the lattice spacing to one, $a=1$, together with the Planck's constant and unit charge $\hbar=q=1$, where energy units will be expressed in terms of $J=1$. 
The BZ is defined by the reciprocal lattice vectors, $\bm{b}_1=2\pi\hat{x}-\frac{2\pi}{\sqrt3}\hat{y}$ and $\bm{b}_2=\frac{4\pi}{\sqrt3}\hat{y}$, which, in the absence of sublattice offset ($\Delta_{\alpha}=0$), harbors two dispersive bands with linear band touchings in between at $K$ and $K'$ as shown in Fig.~\ref{fig1}(b,c), where empty/filled triangles (and other markers) represent opposite charges in a given gap. Crucially, the completely flat third band with a quadratic band touching at $\Gamma$ point carries a patch Euler class $\chi_{23}=1$ between bands 2 and 3, owing to the obstruction to annihilate this double node.

We imagine a periodic modulation in the two-dimensional plane of the lattice, making the Hamiltonian (\ref{eq:Hamiltonian}) time dependent, whose discrete translation symmetry can be restored by going to the frame comoving with the lattice via a gauge transformation~\cite{Eckardt17_RMP,GoldmanDalibard14_PRX}. Eventually, for the (effective) vector potential $\boldsymbol{A}(t)=A_x\cos(\omega t)\hat{x}-A_y\cos(\omega t+\varphi)\hat{y}$, with $\varphi$ controlling the polarization, the crystal momentum gets modified as $\bk\rightarrow\bk+\boldsymbol{A}(t)$ via the minimal coupling~\cite{DuFiete17_PRB}. In the following, we will focus on linear driving along $x$-direction with $A_y=0$ as a route to illustrate Floquet-induced multi-gap topological phases depicted in Fig.~\ref{fig1}d. With the tuning parameters  $\omega$, $A_x$, $\Delta_A$ and $\Delta_C$, where we fix $\Delta_B=-\Delta_A-\Delta_C$ for simplicity, we invoke phase transitions and evaluate the quasienergy spectra numerically together with the change of the Euler class in different gaps, see also Methods.

\begin{figure}
	\centering\includegraphics[width=1\linewidth]{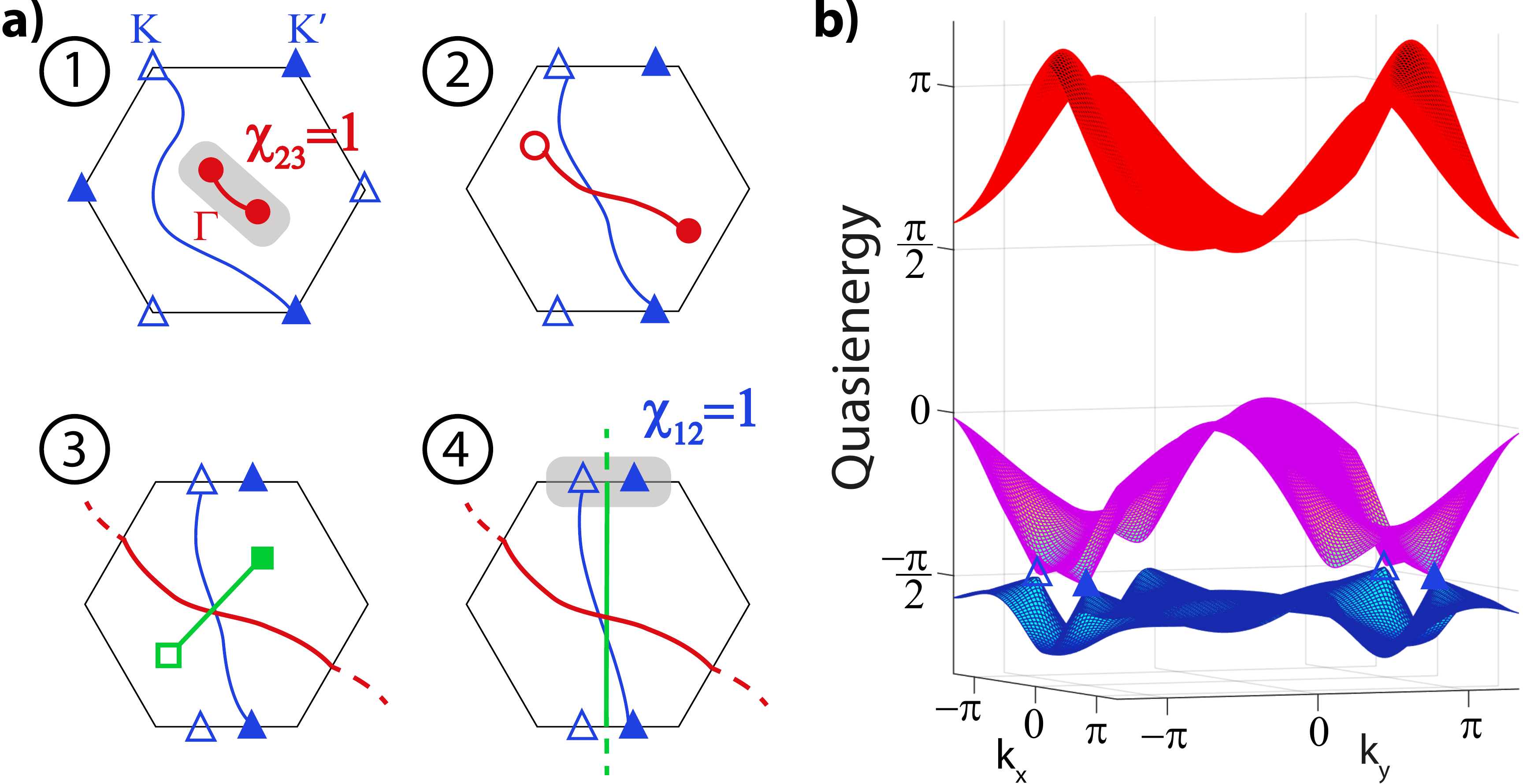}
	\caption{{\bf Realization of an Anomalous Euler phase}. a) Starting from the Kagome model (\ref{eq:Hamiltonian}), linear shaking directly separates the $\Gamma$ nodes (red circles, both filled as they have same charges), giving the Dirac string configuration in step 1. Subsequently, decreasing $\Delta_A$ and $\Delta_C$ (while keeping $\Delta_B=-\Delta_A-\Delta_C$), new nodes (green boxes) in the anomalous gap are created and braided with the existing nodes as shown in steps 2 to 4. The green nodes eventually annihilate across the BZ, leaving behind a Dirac string and obstructing annihilation of the $(K,K')$ nodes (blue triangles). The Euler class has been transferred from the $\Gamma$-node to the $K,K'$ pair, as quantified by final patch Euler classes $\chi_{23}=0$ and $\chi_{12}=1$. b) Quasienergy spectrum of the final phase (4), for $\Delta_A=-2.2, \Delta_C=-1.5$ and $A_x=2,\omega=6$ in units of $J$. }
	\label{fig2}
\end{figure}

{\it Anomalous Euler phase and Floquet-induced braiding --} Using the outlined strategy, we now analyze the first anomalous Euler phase summarized in Fig.~\ref{fig2}a, where the Euler patch class is transferred between different gaps most crucially by braiding through the Dirac string (via the anomalous band nodes) in the anomalous gap at the FBZ edge.

Specifically, upon linearly driving the Kagome lattice (here we consider with a frequency $\omega=6$ and amplitude $A_x=2$), the $\Gamma$-nodes immediately split due to broken ${\cal C}_6$ symmetry, creating a Dirac string (red) in between them (step 1 in Fig.~\ref{fig2}a). We then instigate non-Abelian braiding processes by decreasing the sublattice offsets $\Delta_A$ and $\Delta_C$, allowing us to annihilate the $\Gamma$-nodes across the BZ by flipping its charge upon crossing the blue Dirac string of the ($K,K'$) nodes. The Floquet nature comes into full play as a new pair of nodes (green squares) residing in the $\pi$-gap (i.e.~gap 3) in between the top and bottom bands over the FBZ edge are created and separated with the green Dirac string connecting them, which still carry opposite (empty/filled) charges after crossing two Dirac strings (red, blue) of the adjacent gaps. The last step witnesses the annihilation of the anomalous (green) nodes across the BZ, ensuring their Dirac string is left behind in the middle of $K$ and $K'$. As a result, these nodes in gap 1 now have finite patch Euler class $\chi_{12}=1$ and are thus obstructed to annihilate, which we confirm by evaluating the Euler form for the final stage given in Fig.~\ref{fig2}b for $\Delta_A=-2.2,\Delta_C=-1.5$ (see Methods for details). Consequently, this Euler phase in which the non-trivial patch Euler class has been transferred from the $\Gamma$ to $(K,K')$ nodes is indeed anomalous and does not have a static counterpart. These insights can in fact be corroborated by contrasting with the static system at the same offset potentials, which has trivial Euler class for all subsets of bands where $K$ nodes can annihilate.


We point out that our braiding perspective also shines light on more conventional cases of dynamically inverting band spectra~\cite{Eckardt17_RMP,GoldmanDalibard14_PRX,DuFiete17_PRB}. It is known that under periodic modulation, the tunneling amplitudes get renormalized by Bessel functions~\cite{Eckardt17_RMP}, with the leading order captured by the zeroth-order Bessel function. 
Consequently, tunneling amplitudes can get dynamically frustrated (see e.g.~Ref.\cite{Lignier07_PRL_dynFrustr} for experimental demonstration) or change their sign depending on the ratio of driving strength and frequency~\cite{DuFiete17_PRB}. 
For the Kagome system, inversion results in the  flat band now being the lowest band with the sign change of $J$. Our analysis indeed reveals that this naturally involves braiding of the nodes in gaps 1 and 2 upon increasing the driving amplitude. We detail this in Methods and here focus more on the phases that involve the anomalous gap at the FBZ edge.

\begin{figure}
	\centering\includegraphics[width=.95\linewidth]{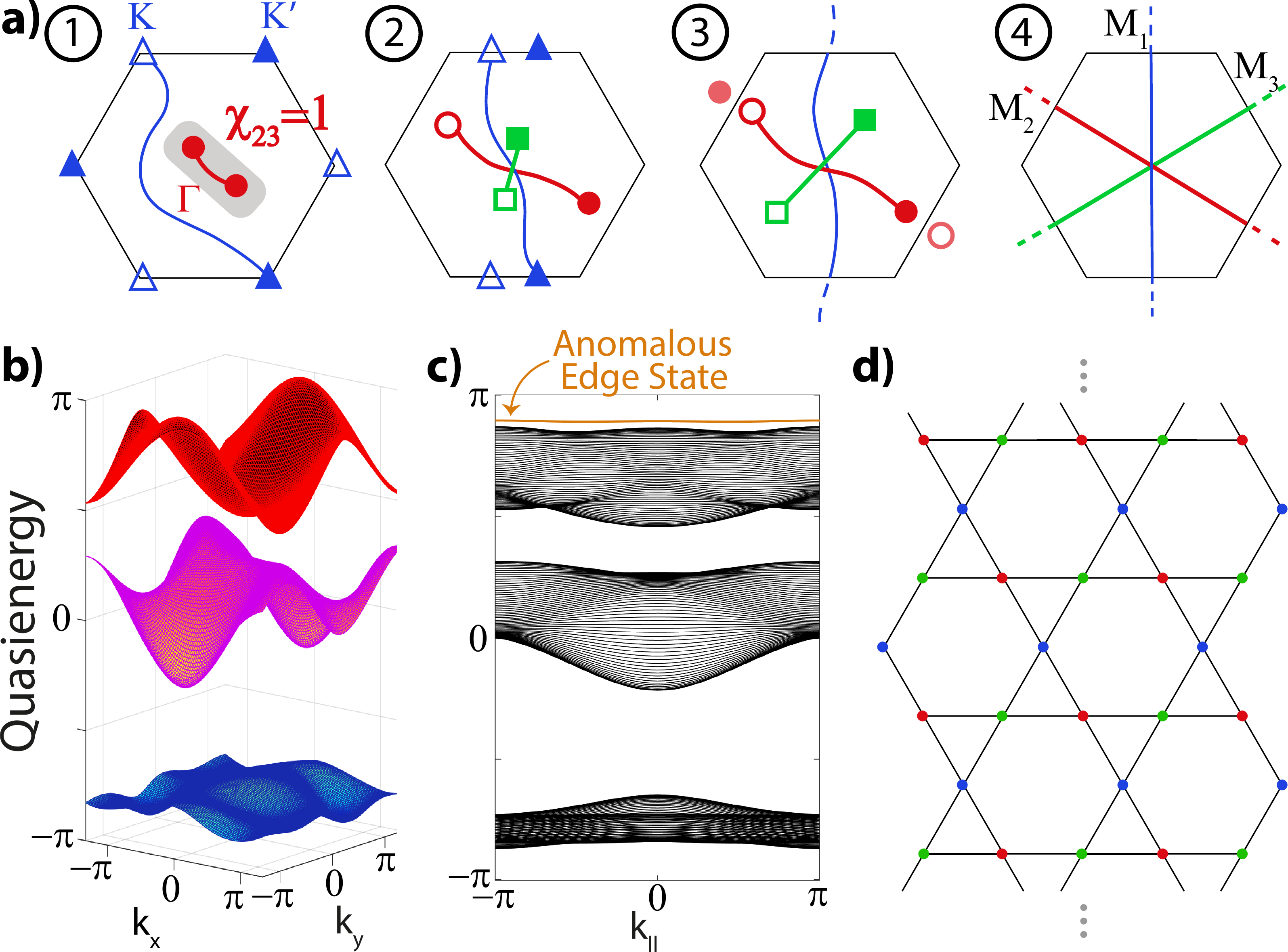}
	\caption{{\bf Realization of an Anomalous Dirac string phase}. Linearly driven Kagome model, with $\omega=6$, $A_x=2$. a) The multi-gap topological configuration evolves as shown in steps $1-4$ upon decreasing $\Delta_C$. The final stage entails the ADS phase, a new fully gapped phase with a Dirac string in each gap, including the anomalous one. b)  Quasienergy spectrum of the ADS phase for $\Delta_A=0, \Delta_C=-\Delta_B=-3$.  c) The extra string of the anomalous gap signals an extra accumulation of $\pi$-Berry phase and thus an unconventional edge spectrum indicated by the relevant Zak phases, for the ribbon geometry depicted in (d). This anomalous edge state acts as a direct observable of the ADS phase. }
	\label{fig3}
\end{figure}

{\it Anomalous Dirac string phase--} We showcase that novel gapped anomalous 
topological phases may arise by an interplay of multi-gap topological principles and the periodicity of the FBZ. Fig.~\ref{fig3} displays an example of a new ``anomalous Dirac string'' (ADS) phase which we obtain by driving with frequency $\omega=6$ and amplitude $A_x=2$ as before while now decreasing only the offset parameter $\Delta_C$ ( i.e.~$\Delta_A=0$). We retrieve the familiar process of splitting the stable double $\Gamma$-node into two nodes that move along the $\mathbf{b}_1$-direction towards one of the $M$ points while $(K,K')$ nodes move towards another one. Further decreasing $\Delta_C=-\Delta_B$ to a value of $-3$, the $(K,K')$ nodes meet and annihilate at $M_1$, leaving behind the blue Dirac string, while the $\Gamma$-nodes in the second gap annihilate at $M_2$ point, creating the red Dirac string along the other direction. The anomalous nature of the phase is induced by the processes in the anomalous gap, which shows the creation of a pair of nodes that, crucially, move to the remaining $M_3$ point at which they annihilate and leave behind the third green Dirac string. Consequently, the system ends up in a totally gapped phase with a Dirac string in each gap. 
As in the anomalous Euler phase, we stress that this gapped ADS phase can only exist in out-of-equilibrium Floquet settings with Dirac strings present in all gaps where the anomalous gap at the FBZ edge allows for the band nodes to annihilate at all different $M$ points.


{\it Edge states of anomalous Dirac string phase--} 
Most importantly, by effectively keeping track of the band inversions via the above described evolution, our analysis also conveys a highlight feature of this esoteric phase that is the appearance of anomalous edge states. While Dirac strings are gauge objects, similar to visons in lattice gauge theories, passing through a Dirac string when considering the Berry phase of non-contractible paths over the BZ does indicate a phase accumulation~\cite{Zak2}, meaning that the Zak phase corresponding to that path shifts by $\pi$. Since the ADS phase is realized upon creating and then annihilating band nodes across the BZ to obtain a Dirac string in the anomalous gap, the bands on either side of the FBZ edge acquire an extra $\pi$-Zak phase~\cite{Zak2}. As a result, edge terminations characterized by Zak phases over paths [perpendicular to this edge] that cross this string should display anomalous edge states, similar to anomalous Floquet topological insulators~\cite{Rudner13_PRX}. We confirm this in the ribbon geometry presented in Fig.~\ref{fig3}(c,d), where the anomalous edge state is precisely identified in the spectrum.
Namely, the Dirac string configuration of the ADS phase differs from the trivial atomic limit only by the presence of the Dirac string in the anomalous gap as demonstrated in the Methods due to the shifted Wannier centers. Hence, the anomalous gap hosts edge states that we indeed prove are completely localized, giving a universal signature of the ADS phase.
This also provides a direct route to create phases with edge states in each gap.

{\it Conclusions and Discussion--}
We present for the first time that the recent developments culminating in novel multi-gap topological phases~\cite{bouhon2020geometric}, have anomalous counterparts that can only exist in an out-of-equilibrium setting. Specifically, we show that the anomalous gap, stemming from the time-periodic nature of the Floquet spectrum, can similarly induce non-Abelian braiding via band nodes therein, leading to Floquet-induced Euler phases. 
Apart from these hitherto uncharted out-of-equilibrium braiding processes, we further discover an anomalous Dirac string phase. 
Akin to the characteristic boundary modes of anomalous Floquet insulators which were employed for their identification, this fully gapped ADS phase has an unconventional Dirac string configuration resulting in a distinct edge state spectrum. These new phases thus appeal both for their non-equilibrium as well as multi-gap nature.
Indeed, we stress that multi-gap models can systematically be formulated for a wide range of systems~\cite{bouhon2020geometric}, which, together with the applicability of periodic driving techniques~\cite{Eckardt17_RMP},
furnishes various routes to realize these new anomalous phases. Here, we specifically mention trapped ion systems~\cite{zhao2022observation}, ultracold atoms~\cite{Wintersperger20_NatPhys,Kurn21_arXiv_bandsingularity,LeungStamperKurn20_PRL_kgm,Cooper19_RMP}, metamaterials~\cite{Jiang2021} or Floquet engineering of real materials~\cite{wang2013observation,Threvisan2022}.
We therefore anticipate that our results form a stepping stone for the theoretical investigation of new exotic topologies and their experimental observation.

\section{Acknowledgements} R.J.S. acknowledges funding from a New Investigator Award, EPSRC grant EP/W00187X/1, as well as Trinity College, Cambridge. A.~B. was funded by a Marie-Curie fellowship, grant no. 101025315. F.N.\"U. was funded by the Royal Society under a Newton International Fellowship and the Marie Sk{\l}odowska-Curie programme of the European Commission Grant No 893915.

\clearpage
\newpage
\section{Methods}
\vspace{-0.5cm}


\subsection{Euler class and multi-gap topological characterizations}
To classify wave functions over a Brillouin zone or Floquet bands over a Floquet Brillouin zone from a topological perspective, one is essentially interested in characterizing the describing vector bundle. Although there has been a wealth of results on getting different topological invariants, we focus on new developments that take into account multi-gap conditions~\cite{bouhon2020geometric}.
We recall that oriented real bundles over a base space $B$ admit a characterizing Euler class, being an element of the de Rham cohomolgy. More specifically, integrating the Euler class~\cite{bouhon2019nonabelian, Milnor:1974, Hatcher_2}, over a base space with no boundary results in integer in units of $2\pi$, meaning that that Euler-form integral defines an element of the singular chomology with integer coefficients $H^2(B,\mathbb{Z})$. From a physical perspective, it has been shown that the Euler class of a two-band isolated subspace directly conveys the stability of the nodes in that subspace. When a system enjoys ${\cal C}_2{\cal T}$ or ${\cal P}{\cal T}$ symmetry, band degeneracies in adjacent gaps, that is in the gap situated directly next to the gap under consideration, carry non-Abelian frame charges \cite{Wu1273, BJY_nielsen, bouhon2019nonabelian}. For example, when one considers a three-band system, the reality condition allows for a frame or dreibein interpretation and the band nodes act like $\pi$-vortices, whose frame charges take values in the quaternion group~\cite{bouhon2019nonabelian}, similar to how disclinations in bi-axial nematic phases can carry quaternionin charges \cite{Kamienrmp, volovik2018investigation, Beekman20171}. From a more mathematical point of view this can be directly seen from the fact that the describing Flag variety relates to  $\mathsf{SO}(3)/\mathsf{D}_2$~\cite{bouhon2019nonabelian, bouhon2020geometric, Wu1273}. The orientation of the frame or dreibein needs to be fixed~\cite{bouhon2020geometric, Unal_quenched_Euler}, ensuring that multiplying any eigenvector spanning the frame with a minus sign is {\it a priori} a gauge degree of freedom, rendering the 
mentioned $\mathsf{SO}(3)/\mathsf{D}_2$ parametrization. The first homotopy group  $\pi_1[\mathsf{SO}(3)/\mathsf{D}_2] = \mathbb{Q}=\{+1,\pm i, \pm j, \pm k, -1\}$ accordingly reveals the frame charges. The Euler class of the gapped isolated two-band subspace then physically pertains to the stabilty of the nodes, giving a finite value when the frame charges of the nodes do not add to zero~\cite{bouhon2019nonabelian}. 

\begin{figure}
	\centering\includegraphics[width=1\linewidth]{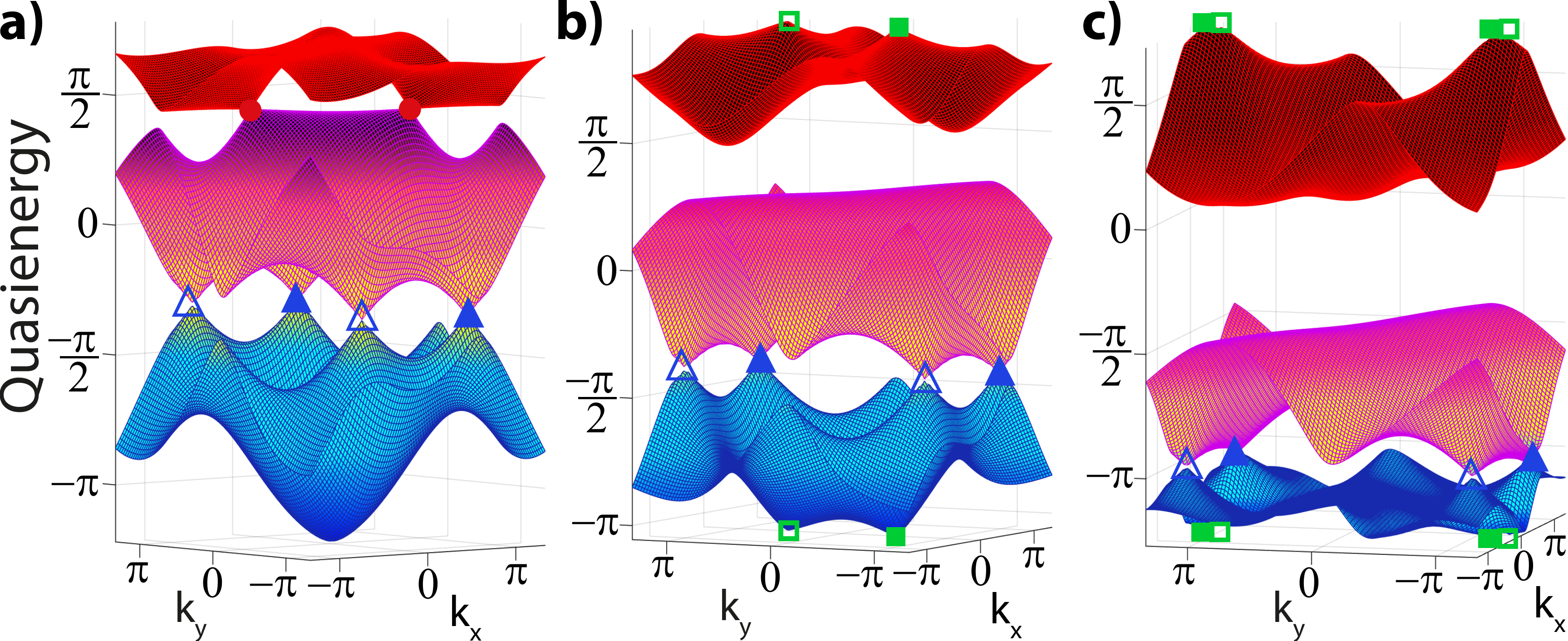}
	\caption{{\bf Floquet spectra of intermediate stages to reach the Anomalous Euler phase.} Band structures during the evolution depicted in Fig.~\ref{fig2} show the anomalous braiding process for frequency $\omega=6$ and amplitude $A_x=2$. The eigenvectors directly characterize the Euler class and quantify the braiding. The offset parameters $[\Delta_A,\Delta_C]$ (where $\Delta_B=-\Delta_A-\Delta_C$) are tuned as a) $[-0.5,-.5]$, b) $[-1,-1]$ and c) $[-1.9,-1.5]$, right before the anomalous (green) nodes annihilate to give rise Fig.~\ref{fig2}b. }
	\label{figS1}
\end{figure}

Most interestingly, the Euler class can be refined to a patch Euler class $\chi$, which essentially evaluates the Euler class over a patch in the Brillouin zone, taking into account a boundary term~\cite{bouhon2019nonabelian}. The explicit expression for a two band subspace spanned by bands $\mathcal{B}_n$ and  $\mathcal{B}_{n+1}$ reads
\begin{widetext}
\begin{equation}\label{eq:Eulerpatch}
\chi[\mathcal{D}]_{n,n+1} \equiv \chi[\{\mathcal{B}_n,\mathcal{B}_{n+1}\};\mathcal{D}] = \dfrac{1}{2\pi} \left[\int_{\mathcal{D}} \mathrm{Eu} - \oint_{\partial \mathcal{D}} \mathrm{a} \right] \in \mathbb{Z},
\end{equation}
\end{widetext}
where the Euler 2-form `Eu' is integrated over the patch $\mathcal{D}$ and supplemented with a boundary term that integrates the Euler connection 1-form `$\mathrm{a}$' over the contour $\partial \mathcal{D}$ of the patch. The Euler 2-form in the above expression in terms of the wave function $u_a$ is defined as $\mathrm{Eu} = d \mathrm{a} = d \mathrm{Pf} \mathcal{A}$ with $\mathcal{A}_{ab} = \langle u_a, \boldsymbol{k} \vert d u_b , \boldsymbol{k}\rangle = \boldsymbol{A}_{ab} \cdot d\boldsymbol{k}= \sum_{i=1,2} \langle u_a, \boldsymbol{k} \vert \partial_{k_i} u_b , \boldsymbol{k}\rangle dk_i $ where we have set $A^{i}_{ab} = \langle u_a, \boldsymbol{k} \vert \partial_{k_i} u_b , \boldsymbol{k}\rangle$ ($A^{i}\in\mathsf{SO}(2)$). This gives $\mathrm{Eu} = (\langle \partial_{k_1} u_a, \boldsymbol{k} \vert \partial_{k_2} u_b, \boldsymbol{k} \rangle - \langle \partial_{k_2} u_a, \boldsymbol{k} \vert \partial_{k_1} u_b, \boldsymbol{k} \rangle ) dk_1\wedge dk_2$, where $a, b$  take values in the band indices $n,n+1$~\cite{bouhon2019nonabelian,BJY_nielsen}. In addition, the Euler connection 1-form is given as $\mathrm{a} =  \mathrm{Pf} \boldsymbol{A} \cdot d\boldsymbol{k}$. 

The above characterization assumes a real bundle structure. In other words we need to maintain either ${\cal C}_2{\cal T}$ or ${\cal P}{\cal T}$ symmetry to ensure that the Hamiltonian and wave functions admit a real formulation. Focusing on Floquet systems the constraint to have real eigenvectors $u_{a}$ restricts the type of the periodic modulation. Hence, we focus on linearly polarized driving as they can preserve time-reversal symmetry in most common ${\cal C}_2$-symmetric setups, fulfilling the reality conditions.

We start from the real-space tight-binding Hamiltonian, $H=-J \sum_{jj'}c^{\dagger}_{j}c_{j'}$, on the Kagome lattice, where $\{j,j'\}$ run through all nearest-neighbor pairs. We consider a periodic driving which induces a force, ${\bf F}(t)=F_x\sin(\omega t)\hat{x}+F_y\sin(\omega t+ \varphi)\hat{y}$ on the lattice site $\bf{r}_j$, making the Hamiltonian time dependent, $H(t)=H-\sum_{j}{\bf F}(t){\bf r}_j\,c^{\dagger}_{j}c_{j}$. After performing a gauge transformation, $R=\exp\{-i\sum_{j}\int dt {\bf F}(t)\,{\bf r}_j\,c^{\dagger}_{j}c_{j}\}$, to the frame of the lattice~\cite{Eckardt17_RMP,Cooper19_RMP}, we obtain a simple form of the Hamiltonian in which the tunneling amplitudes are modified with the Peierls substitution $J_{jj'}(t)=J\exp\{-i\int_{\bf{r}_{j'}}^{\bf{r}_j}\bf{A}(t)\cdot d\bf{r} \}$, where we define the effective vector potential $\boldsymbol{A}(t)=A_x\cos(\omega t)\hat{x}-A_y\cos(\omega t+\varphi)\hat{y}$ with $A_i=F_i/\omega$ [the Planck's constant and lattice spacing are set to one for simplicity]. Hence, we control the strength of our periodic modulation by tuning $(A_x,A_y)$ and, in this work, focus on linear driving only along $x$-direction by setting $A_y=0$. Eventually, this modifies the momentum space Hamiltonian in Eq.\eqref{eq:Hamiltonian} under minimal coupling to $H(\bk,t)= -2J \sum_{\alpha\neq\beta}\cos[(\bk+{\bf A}(t))\cdot{\bf d}_{\alpha\beta}] c^{\dagger}_{\alpha}c_{\beta} +\sum_{\alpha}\Delta_{\alpha}c^{\dagger}_{\alpha}c_{\alpha}$~\cite{DuFiete17_PRB}.
We numerically calculate the time evolution operator at the end of one full period of the drive,  $U(\bk,T)=\mathfrak{T}\exp\{-i\int_0^T H(\bk,t)dt\}$ and evaluate the quasienergy $\varepsilon(\bk)$ as its phase eigenvalues, $e^{-i\varepsilon(\bk)T}$, which can be defined only up to modulo $2\pi$.

\begin{figure}
	\centering\includegraphics[width=.95\linewidth]{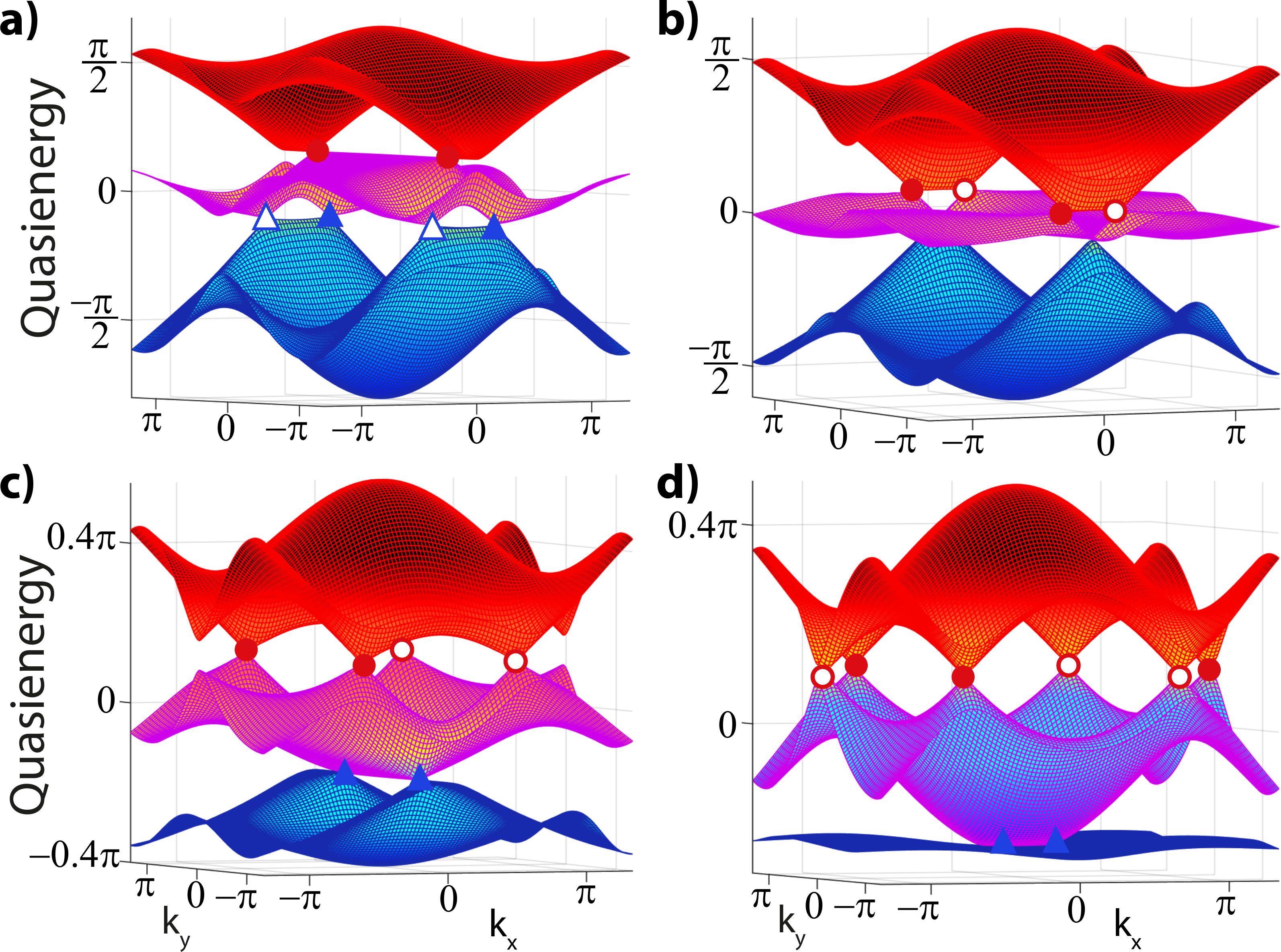}
	\caption{{\bf Floquet non-Abelian braiding and inverted spectrum.} Starting from the static Kagome model, Euler charge is transferred from gap 2 to gap 1 by increasing the driving strength at fixed frequency $\omega=6$ and vanishing sublattice offsets $\Delta_A=\Delta_B=\Delta_C=0$. The driving strength $A_x$ is varied to take values a) 4, b) 5, c) 6 and d) 7.}
	\label{figS6}
\end{figure}

\subsection{Characterizing the Anomalous Euler phase}
With the above characterization of Euler class and multi-gap topology in terms of wave functions, we may directly analyze band spaces of Floquet systems. We note that in fact multi-gap conditions arise naturally as there is no fixed band ordering since the Floquet spectrum repeats itself in multiples of the driving frequency.

Departing from the static regime~\cite{Jiang2021}, we follow the progression of the bands as we tune the parameters of the system, see Fig.~\ref{fig2} of the main text. This achieves a braiding using the nodes of the anomalous gap over the Floquet Brillouin zone, which can be quantified using the patch Euler class of Eq.~\ref{eq:Eulerpatch}, profiting from the discussed real gauge. For completeness, we demonstrate the evolution and the braiding of the band nodes in Fig.~\ref{figS1} which also includes the anomalous band nodes (green) between bands 1 and 3. We confirm these braiding processes by numerically calculating the patch Euler class of the final anomalous Euler phase given in Fig.~\ref{fig2}.

It is instructive to further underpin the truly anomalous nature of the described Euler phase by comparing to its static counterpart with the same sublattice potentials. We stress that in this case the band structure shows a trivial multi-gap topological configuration. Indeed, the nodes between the first and second bands can be gapped, which is corroborated by a calculation of the Euler patch class that is trivial for any patch for the nodal intermediate regions until the band structure is fully gapped.

Secondly, we present another example of Floquet-induced braiding in Fig.~\ref{figS6}, which we now attain by tuning the driving strength while keeping the sublattice offsets at zero, hence, also connecting to the case of dynamically inverting the bands. It is known that the tunneling amplitudes can be frustrated~\cite{Lignier07_PRL_dynFrustr} or made to change sign~\cite{DuFiete17_PRB} by tuning the driving strength. This effect can be understood by looking at effective time-independent tunneling amplitudes at leading order, $J_{\text{eff}}=J{\cal J}_0(A)$~\cite{Eckardt17_RMP}, which gets normalized by the zeroth-order Bessel function ${\cal J}_0$ that can indeed vanish or become negative as a function of $A$. Therefore, one can obtain a spectrum with the flat band with the non-trivial Euler patch class node at the bottom for large $A$, which requires to be addressed from a braiding perspective. As discussed in the main text, linearly driving the Kagome lattice breaks ${\cal C}_6$ symmetry and separates the $\Gamma$-nodes, as shown in Fig.~\ref{figS3}a for $A_x=2$. Upon further increasing the driving strength, our analysis reveals that the $\Gamma$ nodes and $K,K'$ nodes move towards the same $M$ point, where the latter are shown to touch in Fig.~\ref{figS6}b for $A_x=5$. However, instead of annihilating each other and gapping out, the system undergoes rearrangement of Dirac strings where now the $\Gamma$ nodes in gap $2$ carry opposite charges while the nodes in gap $1$ are same valued. This thus effectively amounts to a transfer of Euler charge and the inversion of the band spectrum where the flat band is situated at the bottom. Further increasing the driving strength, we indeed observe that these nodes now move along directions perpendicular to their previous movements and arrange themselves according to the inverted spectrum associated to effective tunneling amplitudes with opposite sign. We stress however that the essential dynamics is captured by a multi-gap braiding perspective, addressing also the topological stability of the $(K,K')$ nodes illustrated with filled triangles in Fig.~\ref{figS6}d.


\subsection{Characterization of Anomalous Dirac String phase}
\begin{figure}
	\centering\includegraphics[width=1\linewidth]{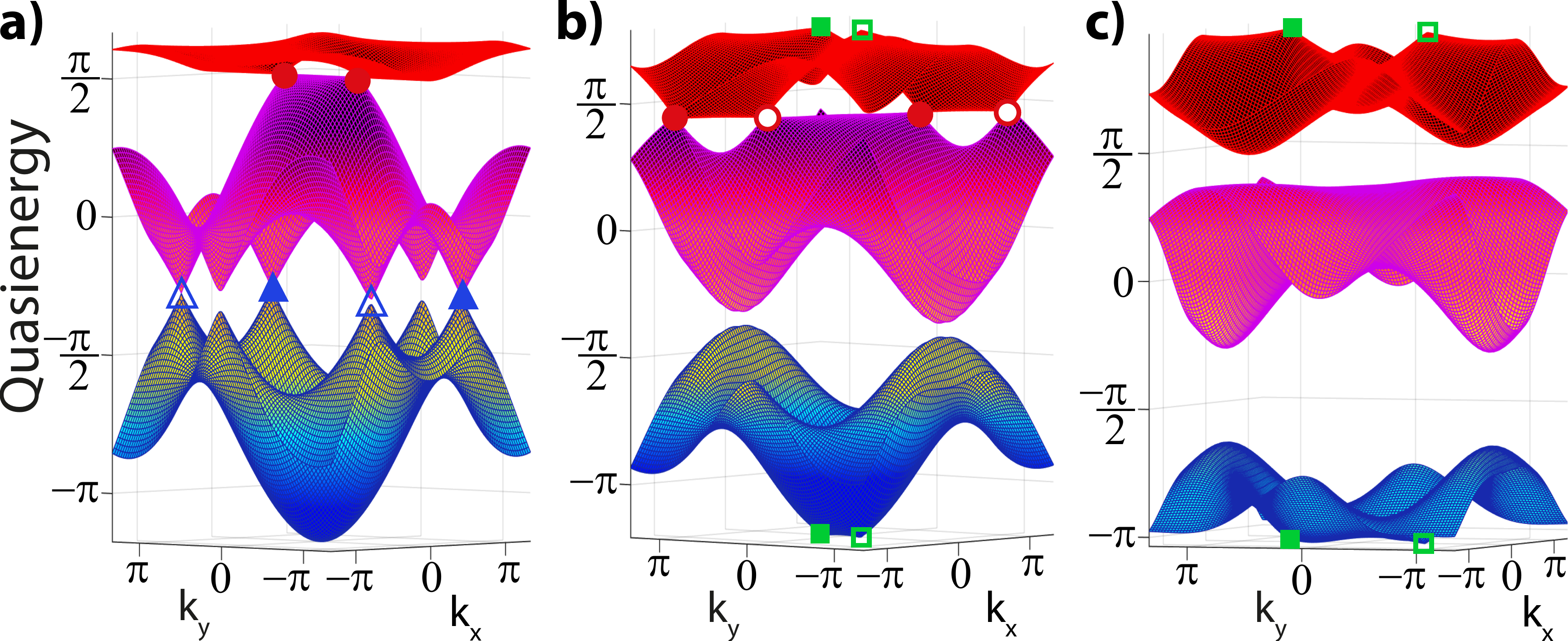}
	\caption{{\bf Floquet band structures of intermediate stages to reach the Anomalous Dirac string phase}. a) Driving [here with frequency $\omega=6$, and amplitude $A_x=2$] splits the double node at the $\Gamma$ point. As detailed in the main text, the ADS phase comes into existence by decreasing $\Delta_C=-\Delta_B$ (at fixed $\Delta_A=0$). Successively, the $(K,K')$ nodes between band 2 and 3  (blue triangles) and the $\Gamma$-nodes (red circles) between bands 1 and 2 annihilate and leave behind Dirac strings shown in Fig.\ref{fig3}a. Similarly, nodes  in the anomalous gap (green boxes) are created and annihilated across the BZ, giving rise to the anomalous Dirac string. The panels show snapshots of the evolution to reach the final stage in Fig.\ref{fig3}b, for a) $\Delta_C=0$, b) $\Delta_C=-1.2$ and c)  $\Delta_C=-2$.}
	\label{figS3}
\end{figure}

Analogous to the Floquet-induced braiding process culminating in the anomalous Euler phase, we can further concretize the Dirac string phase in the outlined model setting, although we stress the generality of these phases. Accordingly, in Fig.~\ref{figS3} we present the band structures of the different stages as shown in Fig.~\ref{fig3}a. A Dirac string physically connects band nodes~\cite{Jiang2021} and thus can be directly tracked by examining the evolution of the bands.  

We can further corroborate the Dirac string configuration by evaluating the Zak phases of the bands, where we recall that a Zak phase is obtained by integrating the Berry phase over a non-contractible path in the Brillouin zone~\cite{Zak1,Zak2}.
As discussed in the main text, when a Dirac string resides between two bands it indicates a phase accumulation of $\pi$ in Berry phase for each crossing of the path with the string. Specifically, turning to the Dirac string configuration of the ADS phase discussed the main text, see also Figs.~\ref{fig3} and \ref{figS5}b),  we have Dirac strings between each band which we will denote as $\text{DS}_{1,2}$, $\text{DS}_{2,3}$ and $\text{DS}_{3,1}$. Here, the subscripts refer to the bands which we number from the lowest to the highest by keeping track of their labeling during the evolution from the static Kagome limit. Hence, $\text{DS}_{1,3}$ corresponds to the Dirac string in the anomalous gap. 

We close the circle of our analysis by relating the Dirac string configuration and the Zak phases of the bands. Given the band inversion processes [Fig.~\ref{figS3}] of the various stages described in the main text that lead to the ADS phase, we obtain the string configuration presented in Fig.~\ref{figS5}b) where the band nodes at each gap are annihilated at a different $M$ point. We can thus readily infer the Zak phase $\gamma_{\mathcal{B}_1}$ of band 1 along the non-contractible path $\mathbf{b}_1$. As this path crosses  $\text{DS}_{1,2}$ and $\text{DS}_{3,1}$, but not $\text{DS}_{2,3}$, we find $\gamma_{\mathcal{B}_1}[\mathbf{b}_1]=0$. This is because band 1 acquires a phase factor $\pi$ from both $\text{DS}_{1,2}$ and $\text{DS}_{3,1}$. Similar reasoning shows that   $\gamma_{\mathcal{B}_2}[\mathbf{b}_1]=\pi$ and $\gamma_{\mathcal{B}_3}[\mathbf{b}_1]=\pi$, 
as these Zak phases get a single $\pi$-contribution from $\text{DS}_{1,2}$ and $\text{DS}_{1,3}$, respectively.
The same procedure can also be repeated for the path along $\mathbf{b}_2$, giving that $(\gamma_{\mathcal{B}_1}[\mathbf{b}_2], \gamma_{\mathcal{B}_2}[\mathbf{b}_2], \gamma_{\mathcal{B}_3}[\mathbf{b}_2])=(\pi,\pi,0)$. Upon numerically calculating the Zak phases of the bands in the ADS regime, we indeed verify these insights and Dirac strings.

\subsection{Edge state counting in Anomalous Dirac String phase}
An important physical consequence that characterizes the anomalous Dirac string (ADS) phase is the appearance of edge states. As we highlighted in the main text, the transition to the ADS phase is characterized by the formation of nodes, that leave behind a Dirac string. While tracking the spectral evolution is effective in characterizing the topological phases, we here like to further quantify the edge state spectra, using these intricate relations between the Zak phases and the various Dirac string configurations. 

In static systems, a Zak phase of $\pi(0)$ indicates the presence (absence) of edge states when the orbitals in real space are centered at the maximally symmetric Wyckoff position~\cite{Zak1,Zak2, solitonsZak}. When the orbitals are ``shifted'', corresponding to the boundary of the unit cell as in the case for our Kagome system, the role of the $0$ and $\pi$-Zak phases is interchanged and a $\pi$-Berry phase corresponds to having no edge states. Essentially the mismatch may be quantified by counting the difference in charges of bands and Wannier centers as formalized by charge anomalies~\cite{solitonsZak}. It is tempting to directly infer from the Zak phases whether an edge termination will give edge states or not in the Floquet setting. Here diligence however has to be taken with the out-of-equilibrium nature of the system. Indeed, due to the time periodicity of the Floquet Brillouin zone, the system is not simply adiabatically connected to the static counterpart, as {\it each} band now relates to gaps above and below. As an example, we consider starting from a static system where the Zak phase configuration of the three bands relevant for a certain edge projection reads $(\gamma_{\mathcal{B}_1}, \gamma_{\mathcal{B}_2}, \gamma_{\mathcal{B}_3})=(0,\pi,\pi)$. Let us further assume that this configuration entails an atomic limit giving no edge states. Upon going to the Floquet counter part, we may consider driving a band inversion in the anomalous gap, inducing a Dirac string and thus a $\pi$ shift in the Zak phase of those bands. This then results in the configuration $(\gamma_{\mathcal{B}_1}, \gamma_{\mathcal{B}_2}, \gamma_{\mathcal{B}_3})=(\pi,\pi,0)$ and an anticipated edge state in the anomalous gap. However, starting from the same initial configuration we may also consider inducing first a band inversion between the two top bands and then the two bottom bands successively. In this scenario, we thus anticipate edge states between bands 1 and 2 and bands 2 and 3. Nonetheless, in this case the Berry phases also amount to  $(\gamma_{\mathcal{B}_1}, \gamma_{\mathcal{B}_2}, \gamma_{\mathcal{B}_3})=(\pi,\pi,0)$, showing the subtly of predicting edge states. Indeed, rather than focusing solely on the Berry phases, one in facts needs to keep track of the Berry phases, Dirac string {\it and} the band evolution.

The above analysis shows that we have to start from a universal description in a well defined limit, a requirement that is unequivocally set by a trivial {\it atomic limit} (AL). To this end, we consider Hamiltonian \eqref{eq:Hamiltonian} with all hopping terms switched off, i.e.~$J=0$, and on site potentials $(\Delta_A,\Delta_B, \Delta_C)=(0,1,-1)$. This evidently realizes a trivially gapped system where the top, middle and bottom band corresponds to the localized wave functions on the $B$, $A$ and $C$ cites, respectively. As the Wannier centers are not localized at the center, this phase does have a Dirac string configuration, as presented in Fig.~\ref{figS5}a). Characterizing the real-space positions of the orbitals $A$, $B$ and $C$ as $\mathbf{r}_{A,B,C}$, a simple calculation then indeed corroborates that
\begin{eqnarray*}
    (\gamma_{\mathcal{B}_1}^{AL}[\mathbf{b}_1], \gamma_{\mathcal{B}_2}^{AL}[\mathbf{b}_1], \gamma_{\mathcal{B}_3}^{AL}[\mathbf{b}_1])&=&(e^{i\mathbf{r}_{C}\cdot\mathbf{b}_1},e^{i\mathbf{r}_{A}\cdot\mathbf{b}_1},e^{i\mathbf{r}_{B}\cdot\mathbf{b}_1})\\&=&(\pi,\pi,0),
\end{eqnarray*}
for the energy-determined band ordering $\mathcal{B}_{1}<\mathcal{B}_{2}< \mathcal{B}_{3}$.
The Zak phases along the other direction are readily verified to amount to $(\gamma_{\mathcal{B}_1}^{AL}[\mathbf{b}_2], \gamma_{\mathcal{B}_2}^{AL}[\mathbf{b}_2], \gamma_{\mathcal{B}_3}^{AL}[\mathbf{b}_2])=(0,\pi,\pi)$. A simple analysis similar to the one presented in the previous subsection then indeed shows that the Dirac string  configuration of the AL phase presented in Fig.~\ref{figS5}a is consistent with these Zak phases.

\begin{figure}
	\centering\includegraphics[width=1\linewidth]{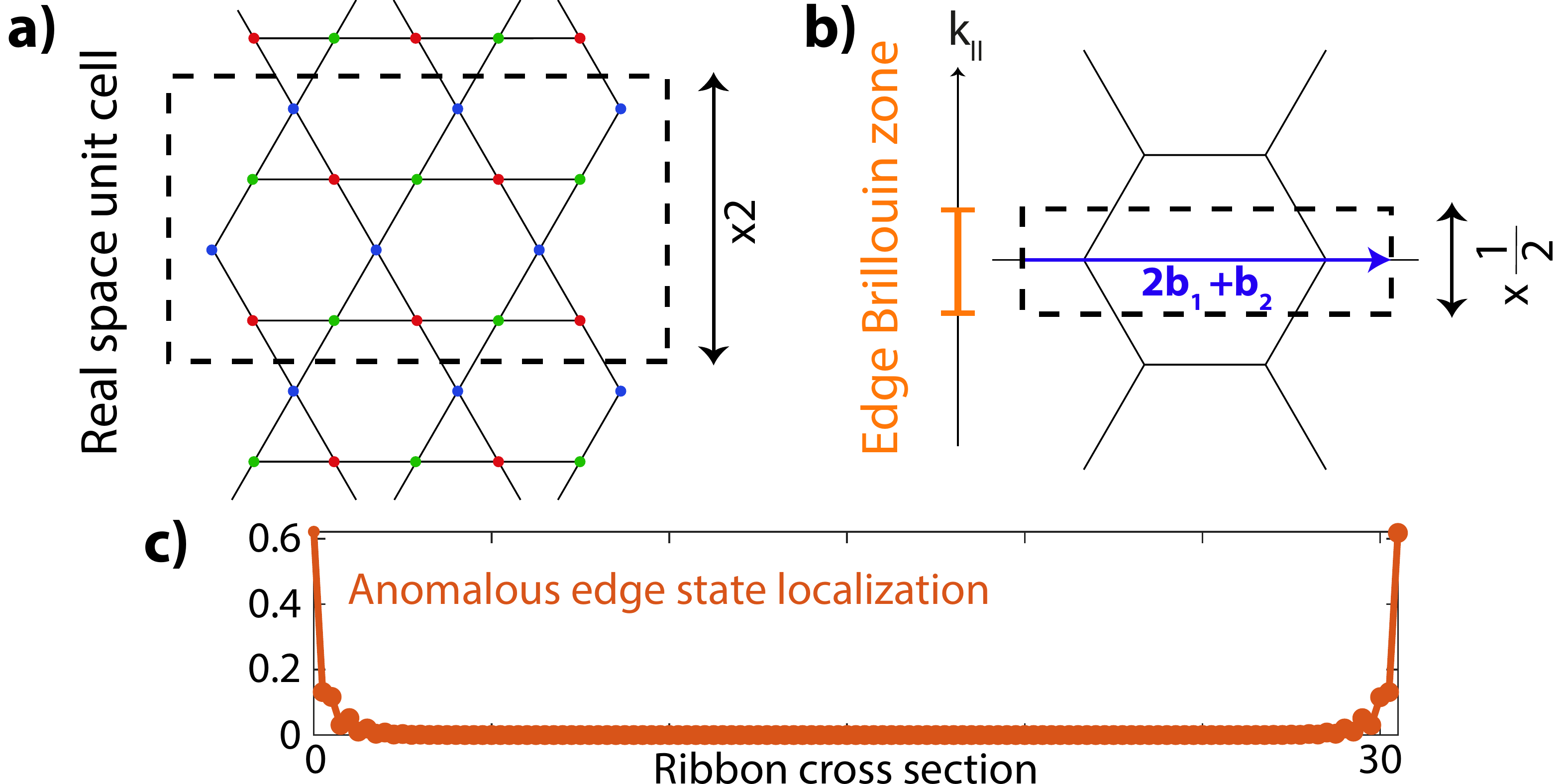}
	\caption{{\bf Edge Brillouin zone determination for `arm chair' cut.} a) The real space ribbon geometry of the main text. The unit cell is doubled along the periodic direction, which results in a halved edge Brillouin zone illustrated in orange (b). The edge states for a band ${\cal B}$ are determined by Zak phases $\gamma_{{\cal B}}[2\mathbf{b}_1+\mathbf{b}_2]$ along paths perpendicular to this edge. c) Real space localization of the two anomalous edge states $\psi_1(x,k_{\parallel})$ and $\psi_2(x,k_{\parallel})$  in the ADS phase as function of $x$ along the ribbon cross section. Due to the difference of the Zak phase configuration in the ADS phase with respect to the atomic limit, these edge states show characteristic real space localization, $\frac{1}{N}\sum_{k_{\parallel}}^N\sum_{y}(|\psi_1(x,y,k_{\parallel})|^2+|\psi_2(x,y,k_{\parallel})|^2)$, where we considered $N=30$ momentum points in the edge BZ and $y$ runs over the sites in the unit cell.}
	\label{figS4}
\end{figure}

\begin{figure}
	\centering\includegraphics[width=1\linewidth]{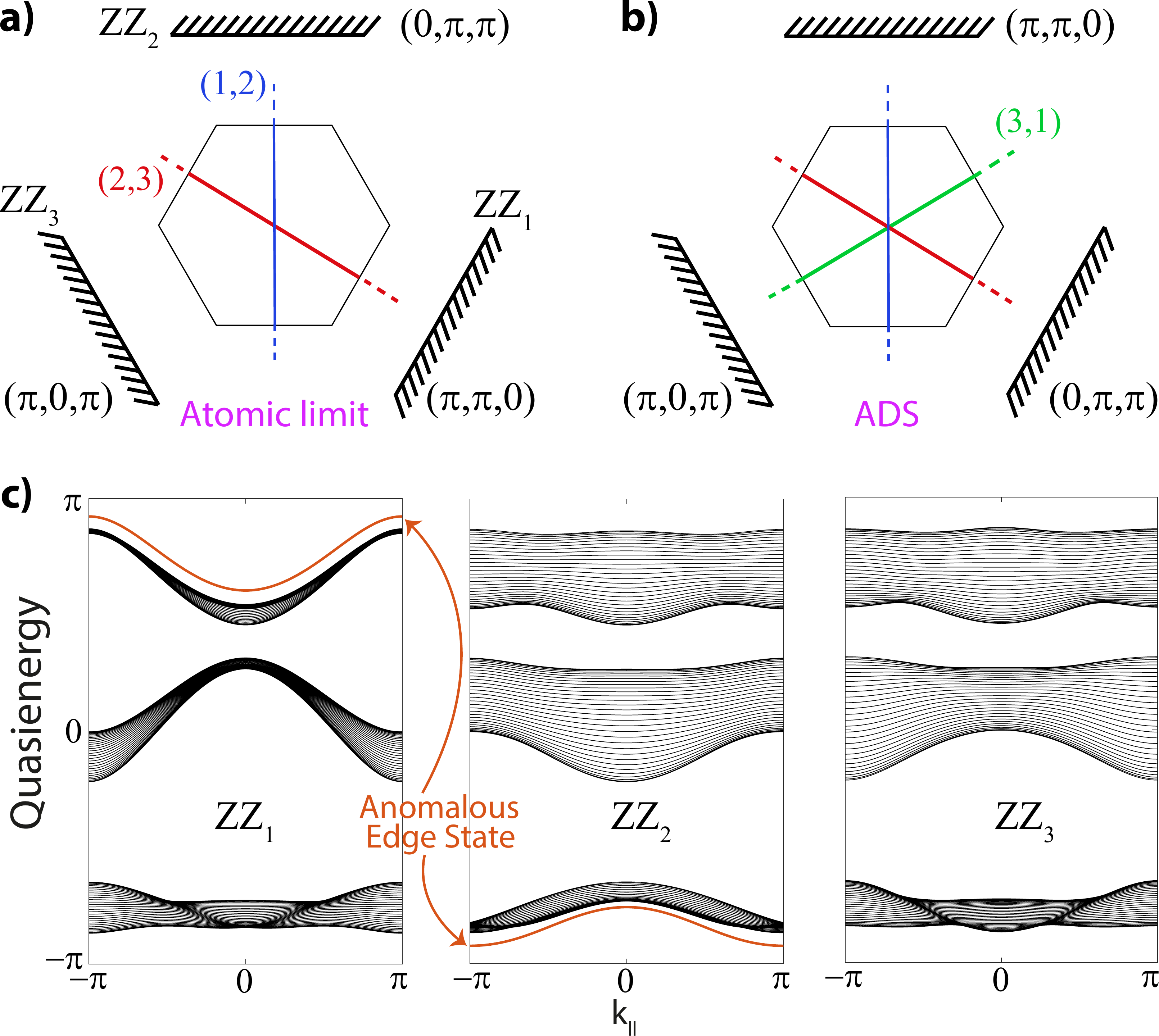}
	\caption{{\bf Zigzag edges and Dirac strings in atomic limit and ADS phase.} a.) Dirac string configuration of the atomic limit defined in text. The three zigzag edges  are characterized by the Zak phases over paths along the perpendicular directions in momentum space. $ZZ_1$, the edge perpendicular to $\mathbf{b}_1$, results in 
	$(\gamma_{\mathcal{B}_1}^{AL}[\mathbf{b}_1], \gamma_{\mathcal{B}_2}^{AL}[\mathbf{b}_1], \gamma_{\mathcal{B}_3}^{AL}[\mathbf{b}_1])=(\pi,\pi,0)$, whereas  $ZZ_2$ and  $ZZ_3$
 are characterized by 		$(\gamma_{\mathcal{B}_1}^{AL}[\mathbf{b}_2], \gamma_{\mathcal{B}_2}^{AL}[\mathbf{b}_2], \gamma_{\mathcal{B}_3}^{AL}[\mathbf{b}_2])=(\pi,0,\pi)$ and 	$(\gamma_{\mathcal{B}_1}^{AL}[\mathbf{v}=-(\mathbf{b}_1-\mathbf{b}_2)], \gamma_{\mathcal{B}_2}^{AL}[\mathbf{v}], \gamma_{\mathcal{B}_3}^{AL}[\mathbf{v}])=(0,\pi,\pi)$, respectively. b) Same edge terminations and relevant Zak phases for the ADS phase, which differs from the atomic limit by the presence of a Dirac string in the anomalous gap. c) Edge state spectra for the zigzag edges in the ADS phase. The $ZZ_1$ and $ZZ_2$ terminations show anomalous edge states due to the difference in Zak phases with respect to the atomic limit, while $ZZ_3$ has no edge states. }
	\label{figS5}
\end{figure}

With the trivial reference state in place, we can now effectively characterize the edge state spectrum for the ``arm chair" termination studied in the main text, in a more concrete manner that corroborates the results of tracking the band inversion processes. In Fig.~\ref{figS4}, we show the ribbon termination and the corresponding momentum space description. The spectrum should be analyzed by the Zak phases that project to the edge Brillouin zone. We observe that for this edge geometry the unit cell is effectively doubled along the periodic direction and hence half as large in reciprocal space. Concretely, the edge sates relate to the Zak phases along the path $2\mathbf{b}_1+\mathbf{b}_2$ perpendicular to the vertical edge. Considering the trivial AL reference phase, it is easy to see that we obtain $(\gamma_{\mathcal{B}_1}^{AL}[2\mathbf{b}_1+\mathbf{b}_2], \gamma_{\mathcal{B}_2}^{AL}[2\mathbf{b}_1+\mathbf{b}_2], \gamma_{\mathcal{B}_3}^{AL}[2\mathbf{b}_1+\mathbf{b}_2])=(0,\pi,\pi)$, as the factor 2 trivializes any contribution from $\mathbf{b}_{1}$, hence ensuring the result is the same as considering the path along $\mathbf{b}_{2}$. Comparing subsequently to the anomalous Dirac string (ADS) phase, we observe that the extra Dirac string induces another $\pi$-phase for the top and bottom band,  $(\gamma_{\mathcal{B}_1}^{ADS}[2\mathbf{b}_1+\mathbf{b}_2], \gamma_{\mathcal{B}_2}^{ADS}[2\mathbf{b}_1+\mathbf{b}_2], \gamma_{\mathcal{B}_3}^{ADS}[2\mathbf{b}_1+\mathbf{b}_2])=(\pi,\pi,0)$. 

We can thus imagine starting from the atomic limit as defined above and induce driving without closing any of the gaps. This phase should then evidently have no edge states. As a next step, the anomalous Dirac string is entered upon a band inversion in the anomalous gap and edge states are anticipated within that gap, as consistent with the observation in Fig.~\ref{fig3} of the main text and quantified via the Zak phase configuration. We reemphasize that the order of processes in the band evolution needs to be taken into account to interpret these indices.

The above presented analysis  can be employed to any type of edge termination. Hence to conclude, we comment on the edge in the three main ``zig zag'' directions demonstrated in Fig.~\ref{figS5}(a,b), where we cut the BZ in three directions that we label as $ZZ_1$, $ZZ_2$ and $ZZ_3$. By applying the similar analysis that we employed in the study of the previous edge termination, we obtain that in the $ZZ_1$-cut the Zak phase configuration reads $(\gamma_{\mathcal{B}_1}^{AL}[\mathbf{b}_1], \gamma_{\mathcal{B}_2}^{AL}[\mathbf{b}_1], \gamma_{\mathcal{B}_3}^{AL}[\mathbf{b}_1])=(\pi,\pi,0)$. Evaluating the configuration in the ADS phase, we get $(\gamma_{\mathcal{B}_1}^{ADS}[\mathbf{b}_1], \gamma_{\mathcal{B}_2}^{ADS}[\mathbf{b}_1], \gamma_{\mathcal{B}_3}^{ADS}[\mathbf{b}_1])=(0,\pi,\pi)$, signaling again that the ADS phase is entered upon a band inversion in the anomalous gap that we verify to be the only gap with edge states in the spectrum for this termination in Fig.~\ref{figS5}c. Turning to the $ZZ_2$-edge, we see that the boundary states are determined by the paths along $\mathbf{b}_2$ and thus result in the same outcomes (Zak phases) as for the ``armchair'' termination above, meaning that we predict edge states in the anomalous gap, which we confirm in Fig.~\ref{figS5}c. Finally, when we consider the $ZZ_3$-edge, the boundary spectrum is determined by paths along the direction $\mathbf{v}=-(\mathbf{b}_1-\mathbf{b}_2)$. We then obtain $(\gamma_{\mathcal{B}_1}^{AL}[\mathbf{v}], \gamma_{\mathcal{B}_2}^{AL}[\mathbf{v}], \gamma_{\mathcal{B}_3}^{AL}[\mathbf{v}])=(\gamma_{\mathcal{B}_1}^{ADS}[\mathbf{v}], \gamma_{\mathcal{B}_2}^{ADS}[\mathbf{v}], \gamma_{\mathcal{B}_3}^{ADS}[\mathbf{v}])=(\pi,0,\pi)$. Consistent with the interpretation that the relevant non-contractible paths in the $\mathbf{v}$-direction do not cross the string of the anomalous gap, no edge states are anticipated for this termination. This is reflected in the similarity of Zak phases for both the ADS phase and AL and corroborated by our numerical results that show no edge states in either of the gaps of our system, see Fig~\ref{figS5}c. We finally note that this explicit evaluation sets the stage for more types of Anomalous Dirac String phases. One may, for example, consider a system that has edge states in the anomalous gap as well as the other gaps. We therefore believe that our results mark an important stepping stone for future pursuits.

\bibliography{references}

\begin{thebibliography}{63}%
\makeatletter
\providecommand \@ifxundefined [1]{%
 \@ifx{#1\undefined}
}%
\providecommand \@ifnum [1]{%
 \ifnum #1\expandafter \@firstoftwo
 \else \expandafter \@secondoftwo
 \fi
}%
\providecommand \@ifx [1]{%
 \ifx #1\expandafter \@firstoftwo
 \else \expandafter \@secondoftwo
 \fi
}%
\providecommand \natexlab [1]{#1}%
\providecommand \enquote  [1]{``#1''}%
\providecommand \bibnamefont  [1]{#1}%
\providecommand \bibfnamefont [1]{#1}%
\providecommand \citenamefont [1]{#1}%
\providecommand \href@noop [0]{\@secondoftwo}%
\providecommand \href [0]{\begingroup \@sanitize@url \@href}%
\providecommand \@href[1]{\@@startlink{#1}\@@href}%
\providecommand \@@href[1]{\endgroup#1\@@endlink}%
\providecommand \@sanitize@url [0]{\catcode `\\12\catcode `\$12\catcode
  `\&12\catcode `\#12\catcode `\^12\catcode `\_12\catcode `\%12\relax}%
\providecommand \@@startlink[1]{}%
\providecommand \@@endlink[0]{}%
\providecommand \url  [0]{\begingroup\@sanitize@url \@url }%
\providecommand \@url [1]{\endgroup\@href {#1}{\urlprefix }}%
\providecommand \urlprefix  [0]{URL }%
\providecommand \Eprint [0]{\href }%
\providecommand \doibase [0]{http://dx.doi.org/}%
\providecommand \selectlanguage [0]{\@gobble}%
\providecommand \bibinfo  [0]{\@secondoftwo}%
\providecommand \bibfield  [0]{\@secondoftwo}%
\providecommand \translation [1]{[#1]}%
\providecommand \BibitemOpen [0]{}%
\providecommand \bibitemStop [0]{}%
\providecommand \bibitemNoStop [0]{.\EOS\space}%
\providecommand \EOS [0]{\spacefactor3000\relax}%
\providecommand \BibitemShut  [1]{\csname bibitem#1\endcsname}%
\let\auto@bib@innerbib\@empty
\bibitem [{\citenamefont {Qi}\ and\ \citenamefont {Zhang}(2011)}]{Rmp1}%
  \BibitemOpen
  \bibfield  {author} {\bibinfo {author} {\bibfnamefont {Xiao-Liang}\
  \bibnamefont {Qi}}\ and\ \bibinfo {author} {\bibfnamefont {Shou-Cheng}\
  \bibnamefont {Zhang}},\ }\bibfield  {title} {\enquote {\bibinfo {title}
  {Topological insulators and superconductors},}\ }\href {\doibase
  10.1103/RevModPhys.83.1057} {\bibfield  {journal} {\bibinfo  {journal} {Rev.
  Mod. Phys.}\ }\textbf {\bibinfo {volume} {83}},\ \bibinfo {pages}
  {1057--1110} (\bibinfo {year} {2011})}\BibitemShut {NoStop}%
\bibitem [{\citenamefont {Hasan}\ and\ \citenamefont {Kane}(2010)}]{Rmp2}%
  \BibitemOpen
  \bibfield  {author} {\bibinfo {author} {\bibfnamefont {M.~Z.}\ \bibnamefont
  {Hasan}}\ and\ \bibinfo {author} {\bibfnamefont {C.~L.}\ \bibnamefont
  {Kane}},\ }\bibfield  {title} {\enquote {\bibinfo {title} {Colloquium:
  {T}opological {I}nsulators},}\ }\href {\doibase 10.1103/RevModPhys.82.3045}
  {\bibfield  {journal} {\bibinfo  {journal} {Rev. Mod. Phys.}\ }\textbf
  {\bibinfo {volume} {82}},\ \bibinfo {pages} {3045--3067} (\bibinfo {year}
  {2010})}\BibitemShut {NoStop}%
\bibitem [{\citenamefont {Armitage}\ \emph {et~al.}(2018)\citenamefont
  {Armitage}, \citenamefont {Mele},\ and\ \citenamefont
  {Vishwanath}}]{Weylrmp}%
  \BibitemOpen
  \bibfield  {author} {\bibinfo {author} {\bibfnamefont {N.~P.}\ \bibnamefont
  {Armitage}}, \bibinfo {author} {\bibfnamefont {E.~J.}\ \bibnamefont {Mele}},
  \ and\ \bibinfo {author} {\bibfnamefont {Ashvin}\ \bibnamefont
  {Vishwanath}},\ }\bibfield  {title} {\enquote {\bibinfo {title} {Weyl and
  dirac semimetals in three-dimensional solids},}\ }\href {\doibase
  10.1103/RevModPhys.90.015001} {\bibfield  {journal} {\bibinfo  {journal}
  {Rev. Mod. Phys.}\ }\textbf {\bibinfo {volume} {90}},\ \bibinfo {pages}
  {015001} (\bibinfo {year} {2018})}\BibitemShut {NoStop}%
\bibitem [{\citenamefont {Fu}(2011)}]{clas1}%
  \BibitemOpen
  \bibfield  {author} {\bibinfo {author} {\bibfnamefont {Liang}\ \bibnamefont
  {Fu}},\ }\bibfield  {title} {\enquote {\bibinfo {title} {Topological
  {C}rystalline {I}nsulators},}\ }\href {\doibase
  10.1103/PhysRevLett.106.106802} {\bibfield  {journal} {\bibinfo  {journal}
  {Phys. Rev. Lett.}\ }\textbf {\bibinfo {volume} {106}},\ \bibinfo {pages}
  {106802} (\bibinfo {year} {2011})}\BibitemShut {NoStop}%
\bibitem [{\citenamefont {Slager}\ \emph {et~al.}(2012)\citenamefont {Slager},
  \citenamefont {Mesaros}, \citenamefont {Juri{\v c}i{\'c}},\ and\
  \citenamefont {Zaanen}}]{clas2}%
  \BibitemOpen
  \bibfield  {author} {\bibinfo {author} {\bibfnamefont {Robert-Jan}\
  \bibnamefont {Slager}}, \bibinfo {author} {\bibfnamefont {Andrej}\
  \bibnamefont {Mesaros}}, \bibinfo {author} {\bibfnamefont {Vladimir}\
  \bibnamefont {Juri{\v c}i{\'c}}}, \ and\ \bibinfo {author} {\bibfnamefont
  {Jan}\ \bibnamefont {Zaanen}},\ }\bibfield  {title} {\enquote {\bibinfo
  {title} {The space group classification of topological band-insulators},}\
  }\href {http://dx.doi.org/10.1038/nphys2513} {\bibfield  {journal} {\bibinfo
  {journal} {Nat. Phys.}\ }\textbf {\bibinfo {volume} {9}},\ \bibinfo {pages}
  {98} (\bibinfo {year} {2012})}\BibitemShut {NoStop}%
\bibitem [{\citenamefont {Soluyanov}\ and\ \citenamefont
  {Vanderbilt}(2012)}]{Vanderbilt_smooth_gauge}%
  \BibitemOpen
  \bibfield  {author} {\bibinfo {author} {\bibfnamefont {Alexey~A.}\
  \bibnamefont {Soluyanov}}\ and\ \bibinfo {author} {\bibfnamefont {David}\
  \bibnamefont {Vanderbilt}},\ }\bibfield  {title} {\enquote {\bibinfo {title}
  {Smooth gauge for topological insulators},}\ }\href {\doibase
  10.1103/PhysRevB.85.115415} {\bibfield  {journal} {\bibinfo  {journal} {Phys.
  Rev. B}\ }\textbf {\bibinfo {volume} {85}},\ \bibinfo {pages} {115415}
  (\bibinfo {year} {2012})}\BibitemShut {NoStop}%
\bibitem [{\citenamefont {Cornfeld}\ and\ \citenamefont
  {Carmeli}(2021)}]{Cornfeld_2021}%
  \BibitemOpen
  \bibfield  {author} {\bibinfo {author} {\bibfnamefont {Eyal}\ \bibnamefont
  {Cornfeld}}\ and\ \bibinfo {author} {\bibfnamefont {Shachar}\ \bibnamefont
  {Carmeli}},\ }\bibfield  {title} {\enquote {\bibinfo {title} {Tenfold
  topology of crystals: Unified classification of crystalline topological
  insulators and superconductors},}\ }\href {\doibase
  10.1103/PhysRevResearch.3.013052} {\bibfield  {journal} {\bibinfo  {journal}
  {Phys. Rev. Research}\ }\textbf {\bibinfo {volume} {3}},\ \bibinfo {pages}
  {013052} (\bibinfo {year} {2021})}\BibitemShut {NoStop}%
\bibitem [{\citenamefont {Turner}\ \emph {et~al.}(2012)\citenamefont {Turner},
  \citenamefont {Zhang}, \citenamefont {Mong},\ and\ \citenamefont
  {Vishwanath}}]{InvTIVish}%
  \BibitemOpen
  \bibfield  {author} {\bibinfo {author} {\bibfnamefont {Ari~M.}\ \bibnamefont
  {Turner}}, \bibinfo {author} {\bibfnamefont {Yi}~\bibnamefont {Zhang}},
  \bibinfo {author} {\bibfnamefont {Roger S.~K.}\ \bibnamefont {Mong}}, \ and\
  \bibinfo {author} {\bibfnamefont {Ashvin}\ \bibnamefont {Vishwanath}},\
  }\bibfield  {title} {\enquote {\bibinfo {title} {Quantized response and
  topology of magnetic insulators with inversion symmetry},}\ }\href {\doibase
  10.1103/PhysRevB.85.165120} {\bibfield  {journal} {\bibinfo  {journal} {Phys.
  Rev. B}\ }\textbf {\bibinfo {volume} {85}},\ \bibinfo {pages} {165120}
  (\bibinfo {year} {2012})}\BibitemShut {NoStop}%
\bibitem [{\citenamefont {Shiozaki}\ and\ \citenamefont
  {Sato}(2014)}]{Shiozaki14}%
  \BibitemOpen
  \bibfield  {author} {\bibinfo {author} {\bibfnamefont {Ken}\ \bibnamefont
  {Shiozaki}}\ and\ \bibinfo {author} {\bibfnamefont {Masatoshi}\ \bibnamefont
  {Sato}},\ }\bibfield  {title} {\enquote {\bibinfo {title} {Topology of
  crystalline insulators and superconductors},}\ }\href {\doibase
  10.1103/PhysRevB.90.165114} {\bibfield  {journal} {\bibinfo  {journal} {Phys.
  Rev. B}\ }\textbf {\bibinfo {volume} {90}},\ \bibinfo {pages} {165114}
  (\bibinfo {year} {2014})}\BibitemShut {NoStop}%
\bibitem [{\citenamefont {Ahn}\ and\ \citenamefont {Yang}(2019)}]{Ahn2019}%
  \BibitemOpen
  \bibfield  {author} {\bibinfo {author} {\bibfnamefont {Junyeong}\
  \bibnamefont {Ahn}}\ and\ \bibinfo {author} {\bibfnamefont {Bohm-Jung}\
  \bibnamefont {Yang}},\ }\bibfield  {title} {\enquote {\bibinfo {title}
  {Symmetry representation approach to topological invariants in
  ${C}_{2z}{T}$-symmetric systems},}\ }\href {\doibase
  10.1103/PhysRevB.99.235125} {\bibfield  {journal} {\bibinfo  {journal} {Phys.
  Rev. B}\ }\textbf {\bibinfo {volume} {99}},\ \bibinfo {pages} {235125}
  (\bibinfo {year} {2019})}\BibitemShut {NoStop}%
\bibitem [{\citenamefont {Alexandradinata}\ and\ \citenamefont
  {Bernevig}(2016)}]{Alex_BerryPhase}%
  \BibitemOpen
  \bibfield  {author} {\bibinfo {author} {\bibfnamefont {A.}~\bibnamefont
  {Alexandradinata}}\ and\ \bibinfo {author} {\bibfnamefont {B.~Andrei}\
  \bibnamefont {Bernevig}},\ }\bibfield  {title} {\enquote {\bibinfo {title}
  {Berry-phase description of topological crystalline insulators},}\ }\href
  {\doibase 10.1103/PhysRevB.93.205104} {\bibfield  {journal} {\bibinfo
  {journal} {Phys. Rev. B}\ }\textbf {\bibinfo {volume} {93}},\ \bibinfo
  {pages} {205104} (\bibinfo {year} {2016})}\BibitemShut {NoStop}%
\bibitem [{\citenamefont {Kruthoff}\ \emph {et~al.}(2017)\citenamefont
  {Kruthoff}, \citenamefont {de~Boer}, \citenamefont {van Wezel}, \citenamefont
  {Kane},\ and\ \citenamefont {Slager}}]{clas3}%
  \BibitemOpen
  \bibfield  {author} {\bibinfo {author} {\bibfnamefont {Jorrit}\ \bibnamefont
  {Kruthoff}}, \bibinfo {author} {\bibfnamefont {Jan}\ \bibnamefont {de~Boer}},
  \bibinfo {author} {\bibfnamefont {Jasper}\ \bibnamefont {van Wezel}},
  \bibinfo {author} {\bibfnamefont {Charles~L.}\ \bibnamefont {Kane}}, \ and\
  \bibinfo {author} {\bibfnamefont {Robert-Jan}\ \bibnamefont {Slager}},\
  }\bibfield  {title} {\enquote {\bibinfo {title} {Topological {C}lassification
  of {C}rystalline {I}nsulators through {B}and {S}tructure {C}ombinatorics},}\
  }\href {\doibase 10.1103/PhysRevX.7.041069} {\bibfield  {journal} {\bibinfo
  {journal} {Phys. Rev. X}\ }\textbf {\bibinfo {volume} {7}},\ \bibinfo {pages}
  {041069} (\bibinfo {year} {2017})}\BibitemShut {NoStop}%
\bibitem [{\citenamefont {Po}\ \emph {et~al.}(2017)\citenamefont {Po},
  \citenamefont {Vishwanath},\ and\ \citenamefont {Watanabe}}]{clas4}%
  \BibitemOpen
  \bibfield  {author} {\bibinfo {author} {\bibfnamefont {Hoi~Chun}\
  \bibnamefont {Po}}, \bibinfo {author} {\bibfnamefont {Ashvin}\ \bibnamefont
  {Vishwanath}}, \ and\ \bibinfo {author} {\bibfnamefont {Haruki}\ \bibnamefont
  {Watanabe}},\ }\bibfield  {title} {\enquote {\bibinfo {title} {Symmetry-based
  indicators of band topology in the 230 space groups},}\ }\href {\doibase
  10.1038/s41467-017-00133-2} {\bibfield  {journal} {\bibinfo  {journal} {Nat.
  Commun.}\ }\textbf {\bibinfo {volume} {8}},\ \bibinfo {pages} {50} (\bibinfo
  {year} {2017})}\BibitemShut {NoStop}%
\bibitem [{\citenamefont {Bradlyn}\ \emph {et~al.}(2017)\citenamefont
  {Bradlyn}, \citenamefont {Elcoro}, \citenamefont {Cano}, \citenamefont
  {Vergniory}, \citenamefont {Wang}, \citenamefont {Felser}, \citenamefont
  {Aroyo},\ and\ \citenamefont {Bernevig}}]{clas5}%
  \BibitemOpen
  \bibfield  {author} {\bibinfo {author} {\bibfnamefont {Barry}\ \bibnamefont
  {Bradlyn}}, \bibinfo {author} {\bibfnamefont {L.}~\bibnamefont {Elcoro}},
  \bibinfo {author} {\bibfnamefont {Jennifer}\ \bibnamefont {Cano}}, \bibinfo
  {author} {\bibfnamefont {M.~G.}\ \bibnamefont {Vergniory}}, \bibinfo {author}
  {\bibfnamefont {Zhijun}\ \bibnamefont {Wang}}, \bibinfo {author}
  {\bibfnamefont {C.}~\bibnamefont {Felser}}, \bibinfo {author} {\bibfnamefont
  {M.~I.}\ \bibnamefont {Aroyo}}, \ and\ \bibinfo {author} {\bibfnamefont
  {B.~Andrei}\ \bibnamefont {Bernevig}},\ }\bibfield  {title} {\enquote
  {\bibinfo {title} {Topological quantum chemistry},}\ }\href
  {http://dx.doi.org/10.1038/nature23268} {\bibfield  {journal} {\bibinfo
  {journal} {Nature}\ }\textbf {\bibinfo {volume} {547}},\ \bibinfo {pages}
  {298} (\bibinfo {year} {2017})}\BibitemShut {NoStop}%
\bibitem [{\citenamefont {Bouhon}\ \emph
  {et~al.}(2020{\natexlab{a}})\citenamefont {Bouhon}, \citenamefont
  {Bzdu\ifmmode~\check{s}\else \v{s}\fi{}ek},\ and\ \citenamefont
  {Slager}}]{bouhon2020geometric}%
  \BibitemOpen
  \bibfield  {author} {\bibinfo {author} {\bibfnamefont {Adrien}\ \bibnamefont
  {Bouhon}}, \bibinfo {author} {\bibfnamefont {Tom\'as}\ \bibnamefont {Bzdu\ifmmode~\check{s}\else
  \v{s}\fi{}ek}}, \ and\ \bibinfo {author} {\bibfnamefont {Robert-Jan}\
  \bibnamefont {Slager}},\ }\bibfield  {title} {\enquote {\bibinfo {title}
  {Geometric approach to fragile topology beyond symmetry indicators},}\ }\href
  {\doibase 10.1103/PhysRevB.102.115135} {\bibfield  {journal} {\bibinfo
  {journal} {Phys. Rev. B}\ }\textbf {\bibinfo {volume} {102}},\ \bibinfo
  {pages} {115135} (\bibinfo {year} {2020}{\natexlab{a}})}\BibitemShut
  {NoStop}%
\bibitem [{\citenamefont {Wu}\ \emph {et~al.}(2019)\citenamefont {Wu},
  \citenamefont {Soluyanov},\ and\ \citenamefont {Bzdu{\v s}ek}}]{Wu1273}%
  \BibitemOpen
  \bibfield  {author} {\bibinfo {author} {\bibfnamefont {QuanSheng}\
  \bibnamefont {Wu}}, \bibinfo {author} {\bibfnamefont {Alexey~A.}\
  \bibnamefont {Soluyanov}}, \ and\ \bibinfo {author} {\bibfnamefont
  {Tom{\'a}{\v s}}\ \bibnamefont {Bzdu{\v s}ek}},\ }\bibfield  {title}
  {\enquote {\bibinfo {title} {Non-{A}belian band topology in noninteracting
  metals},}\ }\href {\doibase 10.1126/science.aau8740} {\bibfield  {journal}
  {\bibinfo  {journal} {Science}\ }\textbf {\bibinfo {volume} {365}},\ \bibinfo
  {pages} {1273--1277} (\bibinfo {year} {2019})}\BibitemShut {NoStop}%
\bibitem [{\citenamefont {Bouhon}\ \emph
  {et~al.}(2020{\natexlab{b}})\citenamefont {Bouhon}, \citenamefont {Wu},
  \citenamefont {Slager}, \citenamefont {Weng}, \citenamefont {Yazyev},\ and\
  \citenamefont {Bzdu{\v s}ek}}]{bouhon2019nonabelian}%
  \BibitemOpen
  \bibfield  {author} {\bibinfo {author} {\bibfnamefont {Adrien}\ \bibnamefont
  {Bouhon}}, \bibinfo {author} {\bibfnamefont {QuanSheng}\ \bibnamefont {Wu}},
  \bibinfo {author} {\bibfnamefont {Robert-Jan}\ \bibnamefont {Slager}},
  \bibinfo {author} {\bibfnamefont {Hongming}\ \bibnamefont {Weng}}, \bibinfo
  {author} {\bibfnamefont {Oleg~V.}\ \bibnamefont {Yazyev}}, \ and\ \bibinfo
  {author} {\bibfnamefont {Tom{\'a}{\v s}}\ \bibnamefont {Bzdu{\v s}ek}},\
  }\bibfield  {title} {\enquote {\bibinfo {title} {Non-abelian reciprocal
  braiding of weyl points and its manifestation in zrte},}\ }\href {\doibase
  10.1038/s41567-020-0967-9} {\bibfield  {journal} {\bibinfo  {journal} {Nature
  Physics}\ }\textbf {\bibinfo {volume} {16}},\ \bibinfo {pages} {1137--1143}
  (\bibinfo {year} {2020}{\natexlab{b}})}\BibitemShut {NoStop}%
\bibitem [{\citenamefont {Ahn}\ \emph {et~al.}(2019)\citenamefont {Ahn},
  \citenamefont {Park},\ and\ \citenamefont {Yang}}]{BJY_nielsen}%
  \BibitemOpen
  \bibfield  {author} {\bibinfo {author} {\bibfnamefont {Junyeong}\
  \bibnamefont {Ahn}}, \bibinfo {author} {\bibfnamefont {Sungjoon}\
  \bibnamefont {Park}}, \ and\ \bibinfo {author} {\bibfnamefont {Bohm-Jung}\
  \bibnamefont {Yang}},\ }\bibfield  {title} {\enquote {\bibinfo {title}
  {Failure of {N}ielsen-{N}inomiya {T}heorem and {F}ragile {T}opology in
  {T}wo-{D}imensional {S}ystems with {S}pace-{T}ime {I}nversion {S}ymmetry:
  {A}pplication to {T}wisted {B}ilayer {G}raphene at {M}agic {A}ngle},}\ }\href
  {\doibase 10.1103/PhysRevX.9.021013} {\bibfield  {journal} {\bibinfo
  {journal} {Phys. Rev. X}\ }\textbf {\bibinfo {volume} {9}},\ \bibinfo {pages}
  {021013} (\bibinfo {year} {2019})}\BibitemShut {NoStop}%
\bibitem [{\citenamefont {Alexander}\ \emph {et~al.}(2012)\citenamefont
  {Alexander}, \citenamefont {Chen}, \citenamefont {Matsumoto},\ and\
  \citenamefont {Kamien}}]{Kamienrmp}%
  \BibitemOpen
  \bibfield  {author} {\bibinfo {author} {\bibfnamefont {Gareth~P.}\
  \bibnamefont {Alexander}}, \bibinfo {author} {\bibfnamefont {Bryan Gin-ge}\
  \bibnamefont {Chen}}, \bibinfo {author} {\bibfnamefont {Elisabetta~A.}\
  \bibnamefont {Matsumoto}}, \ and\ \bibinfo {author} {\bibfnamefont
  {Randall~D.}\ \bibnamefont {Kamien}},\ }\bibfield  {title} {\enquote
  {\bibinfo {title} {{C}olloquium: {D}isclination loops, point defects, and all
  that in nematic liquid crystals},}\ }\href {\doibase
  10.1103/RevModPhys.84.497} {\bibfield  {journal} {\bibinfo  {journal} {Rev.
  Mod. Phys.}\ }\textbf {\bibinfo {volume} {84}},\ \bibinfo {pages} {497--514}
  (\bibinfo {year} {2012})}\BibitemShut {NoStop}%
\bibitem [{\citenamefont {Volovik}\ and\ \citenamefont
  {Mineev}(2018)}]{volovik2018investigation}%
  \BibitemOpen
  \bibfield  {author} {\bibinfo {author} {\bibfnamefont {G.~E.}\ \bibnamefont
  {Volovik}}\ and\ \bibinfo {author} {\bibfnamefont {V.~P.}\ \bibnamefont
  {Mineev}},\ }\bibfield  {title} {\enquote {\bibinfo {title} {Investigation of
  singularities in superfluid {H}e$^3$ in liquid crystals by the homotopic
  topology methods},}\ }in\ \href@noop {} {\emph {\bibinfo {booktitle} {Basic
  Notions Of Condensed Matter Physics}}}\ (\bibinfo  {publisher} {CRC Press},\
  \bibinfo {year} {2018})\ pp.\ \bibinfo {pages} {392--401}\BibitemShut
  {NoStop}%
\bibitem [{\citenamefont {Beekman}\ \emph {et~al.}(2017)\citenamefont
  {Beekman}, \citenamefont {Nissinen}, \citenamefont {Wu}, \citenamefont {Liu},
  \citenamefont {Slager}, \citenamefont {Nussinov}, \citenamefont {Cvetkovic},\
  and\ \citenamefont {Zaanen}}]{Beekman20171}%
  \BibitemOpen
  \bibfield  {author} {\bibinfo {author} {\bibfnamefont {Aron~J.}\ \bibnamefont
  {Beekman}}, \bibinfo {author} {\bibfnamefont {Jaakko}\ \bibnamefont
  {Nissinen}}, \bibinfo {author} {\bibfnamefont {Kai}\ \bibnamefont {Wu}},
  \bibinfo {author} {\bibfnamefont {Ke}~\bibnamefont {Liu}}, \bibinfo {author}
  {\bibfnamefont {Robert-Jan}\ \bibnamefont {Slager}}, \bibinfo {author}
  {\bibfnamefont {Zohar}\ \bibnamefont {Nussinov}}, \bibinfo {author}
  {\bibfnamefont {Vladimir}\ \bibnamefont {Cvetkovic}}, \ and\ \bibinfo
  {author} {\bibfnamefont {Jan}\ \bibnamefont {Zaanen}},\ }\bibfield  {title}
  {\enquote {\bibinfo {title} {Dual gauge field theory of quantum liquid
  crystals in two dimensions},}\ }\href {\doibase
  https://doi.org/10.1016/j.physrep.2017.03.004} {\bibfield  {journal}
  {\bibinfo  {journal} {Phys. Rep.}\ }\textbf {\bibinfo {volume} {683}},\
  \bibinfo {pages} {1 -- 110} (\bibinfo {year} {2017})},\ \bibinfo {note} {dual
  gauge field theory of quantum liquid crystals in two dimensions}\BibitemShut
  {NoStop}%
\bibitem [{\citenamefont {Peng}\ \emph
  {et~al.}(2022{\natexlab{a}})\citenamefont {Peng}, \citenamefont {Bouhon},
  \citenamefont {Monserrat},\ and\ \citenamefont {Slager}}]{Peng2021}%
  \BibitemOpen
  \bibfield  {author} {\bibinfo {author} {\bibfnamefont {Bo}~\bibnamefont
  {Peng}}, \bibinfo {author} {\bibfnamefont {Adrien}\ \bibnamefont {Bouhon}},
  \bibinfo {author} {\bibfnamefont {Bartomeu}\ \bibnamefont {Monserrat}}, \
  and\ \bibinfo {author} {\bibfnamefont {Robert-Jan}\ \bibnamefont {Slager}},\
  }\bibfield  {title} {\enquote {\bibinfo {title} {Phonons as a platform for
  non-abelian braiding and its manifestation in layered silicates},}\ }\href
  {\doibase 10.1038/s41467-022-28046-9} {\bibfield  {journal} {\bibinfo
  {journal} {Nature Communications}\ }\textbf {\bibinfo {volume} {13}},\
  \bibinfo {pages} {423} (\bibinfo {year} {2022}{\natexlab{a}})}\BibitemShut
  {NoStop}%
\bibitem [{\citenamefont {Peng}\ \emph
  {et~al.}(2022{\natexlab{b}})\citenamefont {Peng}, \citenamefont {Bouhon},
  \citenamefont {Slager},\ and\ \citenamefont {Monserrat}}]{peng2022multi}%
  \BibitemOpen
  \bibfield  {author} {\bibinfo {author} {\bibfnamefont {Bo}~\bibnamefont
  {Peng}}, \bibinfo {author} {\bibfnamefont {Adrien}\ \bibnamefont {Bouhon}},
  \bibinfo {author} {\bibfnamefont {Robert-Jan}\ \bibnamefont {Slager}}, \ and\
  \bibinfo {author} {\bibfnamefont {Bartomeu}\ \bibnamefont {Monserrat}},\
  }\bibfield  {title} {\enquote {\bibinfo {title} {Multigap topology and
  non-abelian braiding of phonons from first principles},}\ }\href {\doibase
  10.1103/PhysRevB.105.085115} {\bibfield  {journal} {\bibinfo  {journal}
  {Phys. Rev. B}\ }\textbf {\bibinfo {volume} {105}},\ \bibinfo {pages}
  {085115} (\bibinfo {year} {2022}{\natexlab{b}})}\BibitemShut {NoStop}%
\bibitem [{\citenamefont {Park}\ \emph {et~al.}(2022)\citenamefont {Park},
  \citenamefont {Gao}, \citenamefont {Zhang},\ and\ \citenamefont
  {Oh}}]{park2022nodal}%
  \BibitemOpen
  \bibfield  {author} {\bibinfo {author} {\bibfnamefont {Haedong}\ \bibnamefont
  {Park}}, \bibinfo {author} {\bibfnamefont {Wenlong}\ \bibnamefont {Gao}},
  \bibinfo {author} {\bibfnamefont {Xiao}\ \bibnamefont {Zhang}}, \ and\
  \bibinfo {author} {\bibfnamefont {Sang~Soon}\ \bibnamefont {Oh}},\
  }\href@noop {} {\enquote {\bibinfo {title} {Nodal lines in momentum space:
  topological invariants and recent realizations in photonic and other
  systems},}\ } (\bibinfo {year} {2022}),\ \Eprint
  {http://arxiv.org/abs/2201.06639} {arXiv:2201.06639 [cond-mat.mtrl-sci]}
  \BibitemShut {NoStop}%
\bibitem [{\citenamefont {Park}\ \emph {et~al.}(2021)\citenamefont {Park},
  \citenamefont {Hwang}, \citenamefont {Choi},\ and\ \citenamefont
  {Yang}}]{Park2021}%
  \BibitemOpen
  \bibfield  {author} {\bibinfo {author} {\bibfnamefont {Sungjoon}\
  \bibnamefont {Park}}, \bibinfo {author} {\bibfnamefont {Yoonseok}\
  \bibnamefont {Hwang}}, \bibinfo {author} {\bibfnamefont {Hong~Chul}\
  \bibnamefont {Choi}}, \ and\ \bibinfo {author} {\bibfnamefont {Bohm~Jung}\
  \bibnamefont {Yang}},\ }\bibfield  {title} {\enquote {\bibinfo {title}
  {{Topological acoustic triple point}},}\ }\href {\doibase
  10.1038/s41467-021-27158-y} {\bibfield  {journal} {\bibinfo  {journal}
  {Nature Communications}\ }\textbf {\bibinfo {volume} {12}},\ \bibinfo {pages}
  {1--9} (\bibinfo {year} {2021})}\BibitemShut {NoStop}%
\bibitem [{\citenamefont {Lange}\ \emph {et~al.}(2022)\citenamefont {Lange},
  \citenamefont {Bouhon}, \citenamefont {Monserrat},\ and\ \citenamefont
  {Slager}}]{Lange2022}%
  \BibitemOpen
  \bibfield  {author} {\bibinfo {author} {\bibfnamefont {Gunnar~F.}\
  \bibnamefont {Lange}}, \bibinfo {author} {\bibfnamefont {Adrien}\
  \bibnamefont {Bouhon}}, \bibinfo {author} {\bibfnamefont {Bartomeu}\
  \bibnamefont {Monserrat}}, \ and\ \bibinfo {author} {\bibfnamefont
  {Robert-Jan}\ \bibnamefont {Slager}},\ }\bibfield  {title} {\enquote
  {\bibinfo {title} {Topological continuum charges of acoustic phonons in two
  dimensions and the nambu-goldstone theorem},}\ }\href {\doibase
  10.1103/PhysRevB.105.064301} {\bibfield  {journal} {\bibinfo  {journal}
  {Phys. Rev. B}\ }\textbf {\bibinfo {volume} {105}},\ \bibinfo {pages}
  {064301} (\bibinfo {year} {2022})}\BibitemShut {NoStop}%
\bibitem [{\citenamefont {Chen}\ \emph {et~al.}(2021)\citenamefont {Chen},
  \citenamefont {Bouhon}, \citenamefont {Slager},\ and\ \citenamefont
  {Monserrat}}]{chen2021manipulation}%
  \BibitemOpen
  \bibfield  {author} {\bibinfo {author} {\bibfnamefont {Siyu}\ \bibnamefont
  {Chen}}, \bibinfo {author} {\bibfnamefont {Adrien}\ \bibnamefont {Bouhon}},
  \bibinfo {author} {\bibfnamefont {Robert-Jan}\ \bibnamefont {Slager}}, \ and\
  \bibinfo {author} {\bibfnamefont {Bartomeu}\ \bibnamefont {Monserrat}},\
  }\href@noop {} {\enquote {\bibinfo {title} {Manipulation and braiding of weyl
  nodes using symmetry-constrained phase transitions},}\ } (\bibinfo {year}
  {2021}),\ \Eprint {http://arxiv.org/abs/2108.10330} {arXiv:2108.10330}
  \BibitemShut {NoStop}%
\bibitem [{\citenamefont {Bouhon}\ \emph {et~al.}(2021)\citenamefont {Bouhon},
  \citenamefont {Lange},\ and\ \citenamefont {Slager}}]{magnetic}%
  \BibitemOpen
  \bibfield  {author} {\bibinfo {author} {\bibfnamefont {Adrien}\ \bibnamefont
  {Bouhon}}, \bibinfo {author} {\bibfnamefont {Gunnar~F.}\ \bibnamefont
  {Lange}}, \ and\ \bibinfo {author} {\bibfnamefont {Robert-Jan}\ \bibnamefont
  {Slager}},\ }\bibfield  {title} {\enquote {\bibinfo {title} {Topological
  correspondence between magnetic space group representations and
  subdimensions},}\ }\href {\doibase 10.1103/PhysRevB.103.245127} {\bibfield
  {journal} {\bibinfo  {journal} {Phys. Rev. B}\ }\textbf {\bibinfo {volume}
  {103}},\ \bibinfo {pages} {245127} (\bibinfo {year} {2021})}\BibitemShut
  {NoStop}%
\bibitem [{\citenamefont {Wieder}\ and\ \citenamefont
  {Bernevig}(2018)}]{Wieder_axion}%
  \BibitemOpen
  \bibfield  {author} {\bibinfo {author} {\bibfnamefont {Benjamin~J.}\
  \bibnamefont {Wieder}}\ and\ \bibinfo {author} {\bibfnamefont {B.~Andrei}\
  \bibnamefont {Bernevig}},\ }\href@noop {} {\enquote {\bibinfo {title} {The
  axion insulator as a pump of fragile topology},}\ } (\bibinfo {year}
  {2018}),\ \Eprint {http://arxiv.org/abs/arXiv:1810.02373} {arXiv:1810.02373}
  \BibitemShut {NoStop}%
\bibitem [{\citenamefont {Jiang}\ \emph {et~al.}(2022)\citenamefont {Jiang},
  \citenamefont {Bouhon}, \citenamefont {Wu}, \citenamefont {Kong},
  \citenamefont {Lin}, \citenamefont {Slager},\ and\ \citenamefont
  {Jiang}}]{jiang_meron}%
  \BibitemOpen
  \bibfield  {author} {\bibinfo {author} {\bibfnamefont {Bin}\ \bibnamefont
  {Jiang}}, \bibinfo {author} {\bibfnamefont {Adrien}\ \bibnamefont {Bouhon}},
  \bibinfo {author} {\bibfnamefont {Shi-Qiao}\ \bibnamefont {Wu}}, \bibinfo
  {author} {\bibfnamefont {Ze-Lin}\ \bibnamefont {Kong}}, \bibinfo {author}
  {\bibfnamefont {Zhi-Kang}\ \bibnamefont {Lin}}, \bibinfo {author}
  {\bibfnamefont {Robert-Jan}\ \bibnamefont {Slager}}, \ and\ \bibinfo {author}
  {\bibfnamefont {Jian-Hua}\ \bibnamefont {Jiang}},\ }\bibfield  {title}
  {\enquote {\bibinfo {title} {Experimental observation of meronic topological
  acoustic euler insulators},}\ }\href {\doibase 10.48550/ARXIV.2205.03429} {\
  (\bibinfo {year} {2022}),\ 10.48550/ARXIV.2205.03429}\BibitemShut {NoStop}%
\bibitem [{\citenamefont {Guo}\ \emph {et~al.}(2021)\citenamefont {Guo},
  \citenamefont {Jiang}, \citenamefont {Zhang}, \citenamefont {Zhang},
  \citenamefont {Zhang}, \citenamefont {Yang}, \citenamefont {Zhang},\ and\
  \citenamefont {Chan}}]{Guo1Dexp}%
  \BibitemOpen
  \bibfield  {author} {\bibinfo {author} {\bibfnamefont {Qinghua}\ \bibnamefont
  {Guo}}, \bibinfo {author} {\bibfnamefont {Tianshu}\ \bibnamefont {Jiang}},
  \bibinfo {author} {\bibfnamefont {Ruo-Yang}\ \bibnamefont {Zhang}}, \bibinfo
  {author} {\bibfnamefont {Lei}\ \bibnamefont {Zhang}}, \bibinfo {author}
  {\bibfnamefont {Zhao-Qing}\ \bibnamefont {Zhang}}, \bibinfo {author}
  {\bibfnamefont {Biao}\ \bibnamefont {Yang}}, \bibinfo {author} {\bibfnamefont
  {Shuang}\ \bibnamefont {Zhang}}, \ and\ \bibinfo {author} {\bibfnamefont
  {C.~T.}\ \bibnamefont {Chan}},\ }\bibfield  {title} {\enquote {\bibinfo
  {title} {Experimental observation of non-abelian topological charges and edge
  states},}\ }\href {\doibase 10.1038/s41586-021-03521-3} {\bibfield  {journal}
  {\bibinfo  {journal} {Nature}\ }\textbf {\bibinfo {volume} {594}},\ \bibinfo
  {pages} {195--200} (\bibinfo {year} {2021})}\BibitemShut {NoStop}%
\bibitem [{\citenamefont {Jiang}\ \emph {et~al.}(2021)\citenamefont {Jiang},
  \citenamefont {Bouhon}, \citenamefont {Lin}, \citenamefont {Zhou},
  \citenamefont {Hou}, \citenamefont {Li}, \citenamefont {Slager},\ and\
  \citenamefont {Jiang}}]{Jiang2021}%
  \BibitemOpen
  \bibfield  {author} {\bibinfo {author} {\bibfnamefont {Bin}\ \bibnamefont
  {Jiang}}, \bibinfo {author} {\bibfnamefont {Adrien}\ \bibnamefont {Bouhon}},
  \bibinfo {author} {\bibfnamefont {Zhi-Kang}\ \bibnamefont {Lin}}, \bibinfo
  {author} {\bibfnamefont {Xiaoxi}\ \bibnamefont {Zhou}}, \bibinfo {author}
  {\bibfnamefont {Bo}~\bibnamefont {Hou}}, \bibinfo {author} {\bibfnamefont
  {Feng}\ \bibnamefont {Li}}, \bibinfo {author} {\bibfnamefont {Robert-Jan}\
  \bibnamefont {Slager}}, \ and\ \bibinfo {author} {\bibfnamefont {Jian-Hua}\
  \bibnamefont {Jiang}},\ }\bibfield  {title} {\enquote {\bibinfo {title}
  {Experimental observation of non-abelian topological acoustic semimetals and
  their phase transitions},}\ }\href {\doibase 10.1038/s41567-021-01340-x}
  {\bibfield  {journal} {\bibinfo  {journal} {Nature Physics}\ }\textbf
  {\bibinfo {volume} {17}},\ \bibinfo {pages} {1239--1246} (\bibinfo {year}
  {2021})}\BibitemShut {NoStop}%
\bibitem [{\citenamefont {Qiu}\ \emph {et~al.}(2022)\citenamefont {Qiu},
  \citenamefont {Zhang}, \citenamefont {Liu}, \citenamefont {Fan},
  \citenamefont {Zhang},\ and\ \citenamefont {Qiu}}]{qiu2022minimal}%
  \BibitemOpen
  \bibfield  {author} {\bibinfo {author} {\bibfnamefont {Huahui}\ \bibnamefont
  {Qiu}}, \bibinfo {author} {\bibfnamefont {Qicheng}\ \bibnamefont {Zhang}},
  \bibinfo {author} {\bibfnamefont {Tingzhi}\ \bibnamefont {Liu}}, \bibinfo
  {author} {\bibfnamefont {Xiying}\ \bibnamefont {Fan}}, \bibinfo {author}
  {\bibfnamefont {Fan}\ \bibnamefont {Zhang}}, \ and\ \bibinfo {author}
  {\bibfnamefont {Chunyin}\ \bibnamefont {Qiu}},\ }\bibfield  {title} {\enquote
  {\bibinfo {title} {Minimal non-abelian nodal braiding in ideal
  metamaterials},}\ }\href@noop {} {\  (\bibinfo {year} {2022})},\ \Eprint
  {http://arxiv.org/abs/2202.01467} {arXiv:2202.01467 [cond-mat.other]}
  \BibitemShut {NoStop}%
\bibitem [{\citenamefont {\"Unal}\ \emph {et~al.}(2020)\citenamefont {\"Unal},
  \citenamefont {Bouhon},\ and\ \citenamefont {Slager}}]{Unal_quenched_Euler}%
  \BibitemOpen
  \bibfield  {author} {\bibinfo {author} {\bibfnamefont {F.~Nur}\ \bibnamefont
  {\"Unal}}, \bibinfo {author} {\bibfnamefont {Adrien}\ \bibnamefont {Bouhon}},
  \ and\ \bibinfo {author} {\bibfnamefont {Robert-Jan}\ \bibnamefont
  {Slager}},\ }\bibfield  {title} {\enquote {\bibinfo {title} {Topological
  euler class as a dynamical observable in optical lattices},}\ }\href
  {\doibase 10.1103/PhysRevLett.125.053601} {\bibfield  {journal} {\bibinfo
  {journal} {Phys. Rev. Lett.}\ }\textbf {\bibinfo {volume} {125}},\ \bibinfo
  {pages} {053601} (\bibinfo {year} {2020})}\BibitemShut {NoStop}%
\bibitem [{\citenamefont {Zhao}\ \emph {et~al.}(2022)\citenamefont {Zhao},
  \citenamefont {Yang}, \citenamefont {Jiang}, \citenamefont {Mao},
  \citenamefont {Guo}, \citenamefont {Qiu}, \citenamefont {Wang}, \citenamefont
  {Yao}, \citenamefont {He}, \citenamefont {Zhou}, \citenamefont {Xu},\ and\
  \citenamefont {Duan}}]{zhao2022observation}%
  \BibitemOpen
  \bibfield  {author} {\bibinfo {author} {\bibfnamefont {W.~D.}\ \bibnamefont
  {Zhao}}, \bibinfo {author} {\bibfnamefont {Y.~B.}\ \bibnamefont {Yang}},
  \bibinfo {author} {\bibfnamefont {Y.}~\bibnamefont {Jiang}}, \bibinfo
  {author} {\bibfnamefont {Z.~C.}\ \bibnamefont {Mao}}, \bibinfo {author}
  {\bibfnamefont {W.~X.}\ \bibnamefont {Guo}}, \bibinfo {author} {\bibfnamefont
  {L.~Y.}\ \bibnamefont {Qiu}}, \bibinfo {author} {\bibfnamefont {G.~X.}\
  \bibnamefont {Wang}}, \bibinfo {author} {\bibfnamefont {L.}~\bibnamefont
  {Yao}}, \bibinfo {author} {\bibfnamefont {L.}~\bibnamefont {He}}, \bibinfo
  {author} {\bibfnamefont {Z.~C.}\ \bibnamefont {Zhou}}, \bibinfo {author}
  {\bibfnamefont {Y.}~\bibnamefont {Xu}}, \ and\ \bibinfo {author}
  {\bibfnamefont {L.~M.}\ \bibnamefont {Duan}},\ }\href@noop {} {\enquote
  {\bibinfo {title} {Observation of topological euler insulators with a
  trapped-ion quantum simulator},}\ } (\bibinfo {year} {2022}),\ \Eprint
  {http://arxiv.org/abs/2201.09234} {arXiv:2201.09234 [quant-ph]} \BibitemShut
  {NoStop}%
\bibitem [{\citenamefont {Roy}\ and\ \citenamefont {Harper}(2017)}]{Roy17_PRB}%
  \BibitemOpen
  \bibfield  {author} {\bibinfo {author} {\bibfnamefont {Rahul}\ \bibnamefont
  {Roy}}\ and\ \bibinfo {author} {\bibfnamefont {Fenner}\ \bibnamefont
  {Harper}},\ }\bibfield  {title} {\enquote {\bibinfo {title} {Periodic table
  for {F}loquet topological insulators},}\ }\href {\doibase
  10.1103/PhysRevB.96.155118} {\bibfield  {journal} {\bibinfo  {journal} {Phys.
  Rev. B}\ }\textbf {\bibinfo {volume} {96}},\ \bibinfo {pages} {155118}
  (\bibinfo {year} {2017})}\BibitemShut {NoStop}%
\bibitem [{\citenamefont {Kitagawa}\ \emph {et~al.}(2010)\citenamefont
  {Kitagawa}, \citenamefont {Berg}, \citenamefont {Rudner},\ and\ \citenamefont
  {Demler}}]{Kitagawa10_PRB}%
  \BibitemOpen
  \bibfield  {author} {\bibinfo {author} {\bibfnamefont {Takuya}\ \bibnamefont
  {Kitagawa}}, \bibinfo {author} {\bibfnamefont {Erez}\ \bibnamefont {Berg}},
  \bibinfo {author} {\bibfnamefont {Mark}\ \bibnamefont {Rudner}}, \ and\
  \bibinfo {author} {\bibfnamefont {Eugene}\ \bibnamefont {Demler}},\
  }\bibfield  {title} {\enquote {\bibinfo {title} {Topological characterization
  of periodically driven quantum systems},}\ }\href {\doibase
  10.1103/PhysRevB.82.235114} {\bibfield  {journal} {\bibinfo  {journal} {Phys.
  Rev. B}\ }\textbf {\bibinfo {volume} {82}},\ \bibinfo {pages} {235114}
  (\bibinfo {year} {2010})}\BibitemShut {NoStop}%
\bibitem [{\citenamefont {Rudner}\ \emph {et~al.}(2013)\citenamefont {Rudner},
  \citenamefont {Lindner}, \citenamefont {Berg},\ and\ \citenamefont
  {Levin}}]{Rudner13_PRX}%
  \BibitemOpen
  \bibfield  {author} {\bibinfo {author} {\bibfnamefont {Mark~S.}\ \bibnamefont
  {Rudner}}, \bibinfo {author} {\bibfnamefont {Netanel~H.}\ \bibnamefont
  {Lindner}}, \bibinfo {author} {\bibfnamefont {Erez}\ \bibnamefont {Berg}}, \
  and\ \bibinfo {author} {\bibfnamefont {Michael}\ \bibnamefont {Levin}},\
  }\bibfield  {title} {\enquote {\bibinfo {title} {Anomalous edge states and
  the bulk-edge correspondence for periodically driven two-dimensional
  systems},}\ }\href {\doibase 10.1103/PhysRevX.3.031005} {\bibfield  {journal}
  {\bibinfo  {journal} {Phys. Rev. X}\ }\textbf {\bibinfo {volume} {3}},\
  \bibinfo {pages} {031005} (\bibinfo {year} {2013})}\BibitemShut {NoStop}%
\bibitem [{\citenamefont {\"Unal}\ \emph
  {et~al.}(2019{\natexlab{a}})\citenamefont {\"Unal}, \citenamefont {Eckardt},\
  and\ \citenamefont {Slager}}]{Unal19_PRR}%
  \BibitemOpen
  \bibfield  {author} {\bibinfo {author} {\bibfnamefont {F.~Nur}\ \bibnamefont
  {\"Unal}}, \bibinfo {author} {\bibfnamefont {Andr\'e}\ \bibnamefont
  {Eckardt}}, \ and\ \bibinfo {author} {\bibfnamefont {Robert-Jan}\
  \bibnamefont {Slager}},\ }\bibfield  {title} {\enquote {\bibinfo {title}
  {Hopf characterization of two-dimensional floquet topological insulators},}\
  }\href {\doibase 10.1103/PhysRevResearch.1.022003} {\bibfield  {journal}
  {\bibinfo  {journal} {Phys. Rev. Research}\ }\textbf {\bibinfo {volume}
  {1}},\ \bibinfo {pages} {022003(R)} (\bibinfo {year}
  {2019}{\natexlab{a}})}\BibitemShut {NoStop}%
\bibitem [{\citenamefont {\"Unal}\ \emph {et~al.}(2016)\citenamefont {\"Unal},
  \citenamefont {Mueller},\ and\ \citenamefont {Oktel}}]{Unal16_PRA}%
  \BibitemOpen
  \bibfield  {author} {\bibinfo {author} {\bibfnamefont {F.~Nur}\ \bibnamefont
  {\"Unal}}, \bibinfo {author} {\bibfnamefont {Erich~J.}\ \bibnamefont
  {Mueller}}, \ and\ \bibinfo {author} {\bibfnamefont {M.~\"O.}\ \bibnamefont
  {Oktel}},\ }\bibfield  {title} {\enquote {\bibinfo {title} {Nonequilibrium
  fractional hall response after a topological quench},}\ }\href {\doibase
  10.1103/PhysRevA.94.053604} {\bibfield  {journal} {\bibinfo  {journal} {Phys.
  Rev. A}\ }\textbf {\bibinfo {volume} {94}},\ \bibinfo {pages} {053604}
  (\bibinfo {year} {2016})}\BibitemShut {NoStop}%
\bibitem [{\citenamefont {Wang}\ \emph {et~al.}(2017)\citenamefont {Wang},
  \citenamefont {Zhang}, \citenamefont {Chen}, \citenamefont {Yu},\ and\
  \citenamefont {Zhai}}]{Wangchern_17_PRL}%
  \BibitemOpen
  \bibfield  {author} {\bibinfo {author} {\bibfnamefont {Ce}~\bibnamefont
  {Wang}}, \bibinfo {author} {\bibfnamefont {Pengfei}\ \bibnamefont {Zhang}},
  \bibinfo {author} {\bibfnamefont {Xin}\ \bibnamefont {Chen}}, \bibinfo
  {author} {\bibfnamefont {Jinlong}\ \bibnamefont {Yu}}, \ and\ \bibinfo
  {author} {\bibfnamefont {Hui}\ \bibnamefont {Zhai}},\ }\bibfield  {title}
  {\enquote {\bibinfo {title} {Scheme to measure the topological number of a
  chern insulator from quench dynamics},}\ }\href {\doibase
  10.1103/PhysRevLett.118.185701} {\bibfield  {journal} {\bibinfo  {journal}
  {Phys. Rev. Lett.}\ }\textbf {\bibinfo {volume} {118}},\ \bibinfo {pages}
  {185701} (\bibinfo {year} {2017})}\BibitemShut {NoStop}%
\bibitem [{\citenamefont {Goldman}\ and\ \citenamefont
  {Dalibard}(2014)}]{GoldmanDalibard14_PRX}%
  \BibitemOpen
  \bibfield  {author} {\bibinfo {author} {\bibfnamefont {N.}~\bibnamefont
  {Goldman}}\ and\ \bibinfo {author} {\bibfnamefont {J.}~\bibnamefont
  {Dalibard}},\ }\bibfield  {title} {\enquote {\bibinfo {title} {Periodically
  driven quantum systems: Effective hamiltonians and engineered gauge
  fields},}\ }\href {\doibase 10.1103/PhysRevX.4.031027} {\bibfield  {journal}
  {\bibinfo  {journal} {Phys. Rev. X}\ }\textbf {\bibinfo {volume} {4}},\
  \bibinfo {pages} {031027} (\bibinfo {year} {2014})}\BibitemShut {NoStop}%
\bibitem [{\citenamefont {Eckardt}(2017)}]{Eckardt17_RMP}%
  \BibitemOpen
  \bibfield  {author} {\bibinfo {author} {\bibfnamefont {Andr\'e}\ \bibnamefont
  {Eckardt}},\ }\bibfield  {title} {\enquote {\bibinfo {title} {Colloquium:
  Atomic quantum gases in periodically driven optical lattices},}\ }\href
  {\doibase 10.1103/RevModPhys.89.011004} {\bibfield  {journal} {\bibinfo
  {journal} {Rev. Mod. Phys.}\ }\textbf {\bibinfo {volume} {89}},\ \bibinfo
  {pages} {011004} (\bibinfo {year} {2017})}\BibitemShut {NoStop}%
\bibitem [{\citenamefont {Cooper}\ \emph {et~al.}(2019)\citenamefont {Cooper},
  \citenamefont {Dalibard},\ and\ \citenamefont {Spielman}}]{Cooper19_RMP}%
  \BibitemOpen
  \bibfield  {author} {\bibinfo {author} {\bibfnamefont {N.~R.}\ \bibnamefont
  {Cooper}}, \bibinfo {author} {\bibfnamefont {J.}~\bibnamefont {Dalibard}}, \
  and\ \bibinfo {author} {\bibfnamefont {I.~B.}\ \bibnamefont {Spielman}},\
  }\bibfield  {title} {\enquote {\bibinfo {title} {Topological bands for
  ultracold atoms},}\ }\href {\doibase 10.1103/RevModPhys.91.015005} {\bibfield
   {journal} {\bibinfo  {journal} {Rev. Mod. Phys.}\ }\textbf {\bibinfo
  {volume} {91}},\ \bibinfo {pages} {015005} (\bibinfo {year}
  {2019})}\BibitemShut {NoStop}%
\bibitem [{\citenamefont {Hu}\ and\ \citenamefont {Zhao}(2020)}]{HuZhao20_PRL}%
  \BibitemOpen
  \bibfield  {author} {\bibinfo {author} {\bibfnamefont {Haiping}\ \bibnamefont
  {Hu}}\ and\ \bibinfo {author} {\bibfnamefont {Erhai}\ \bibnamefont {Zhao}},\
  }\bibfield  {title} {\enquote {\bibinfo {title} {Topological invariants for
  quantum quench dynamics from unitary evolution},}\ }\href {\doibase
  10.1103/PhysRevLett.124.160402} {\bibfield  {journal} {\bibinfo  {journal}
  {Phys. Rev. Lett.}\ }\textbf {\bibinfo {volume} {124}},\ \bibinfo {pages}
  {160402} (\bibinfo {year} {2020})}\BibitemShut {NoStop}%
\bibitem [{\citenamefont {Wang}\ \emph {et~al.}(2018)\citenamefont {Wang},
  \citenamefont {\"Unal},\ and\ \citenamefont {Eckardt}}]{WangUnal_18_PRL}%
  \BibitemOpen
  \bibfield  {author} {\bibinfo {author} {\bibfnamefont {Botao}\ \bibnamefont
  {Wang}}, \bibinfo {author} {\bibfnamefont {F.~Nur}\ \bibnamefont {\"Unal}}, \
  and\ \bibinfo {author} {\bibfnamefont {Andr\'e}\ \bibnamefont {Eckardt}},\
  }\bibfield  {title} {\enquote {\bibinfo {title} {Floquet engineering of
  optical solenoids and quantized charge pumping along tailored paths in
  two-dimensional {C}hern insulators},}\ }\href {\doibase
  10.1103/PhysRevLett.120.243602} {\bibfield  {journal} {\bibinfo  {journal}
  {Phys. Rev. Lett.}\ }\textbf {\bibinfo {volume} {120}},\ \bibinfo {pages}
  {243602} (\bibinfo {year} {2018})}\BibitemShut {NoStop}%
\bibitem [{\citenamefont {Ra\ifmmode \check{c}\else
  \v{c}\fi{}i\ifmmode~\bar{u}\else \={u}\fi{}nas}\ \emph
  {et~al.}(2018)\citenamefont {Ra\ifmmode \check{c}\else
  \v{c}\fi{}i\ifmmode~\bar{u}\else \={u}\fi{}nas}, \citenamefont {\"Unal},
  \citenamefont {Anisimovas},\ and\ \citenamefont
  {Eckardt}}]{RaciunasUnal_18_PRA}%
  \BibitemOpen
  \bibfield  {author} {\bibinfo {author} {\bibfnamefont {Mantas}\ \bibnamefont
  {Ra\ifmmode \check{c}\else \v{c}\fi{}i\ifmmode~\bar{u}\else \={u}\fi{}nas}},
  \bibinfo {author} {\bibfnamefont {F.~Nur}\ \bibnamefont {\"Unal}}, \bibinfo
  {author} {\bibfnamefont {Egidijus}\ \bibnamefont {Anisimovas}}, \ and\
  \bibinfo {author} {\bibfnamefont {Andr\'e}\ \bibnamefont {Eckardt}},\
  }\bibfield  {title} {\enquote {\bibinfo {title} {Creating, probing, and
  manipulating fractionally charged excitations of fractional chern insulators
  in optical lattices},}\ }\href {\doibase 10.1103/PhysRevA.98.063621}
  {\bibfield  {journal} {\bibinfo  {journal} {Phys. Rev. A}\ }\textbf {\bibinfo
  {volume} {98}},\ \bibinfo {pages} {063621} (\bibinfo {year}
  {2018})}\BibitemShut {NoStop}%
\bibitem [{\citenamefont {Wintersperger}\ \emph {et~al.}(2020)\citenamefont
  {Wintersperger}, \citenamefont {Braun}, \citenamefont {\"Unal}, \citenamefont
  {Eckardt}, \citenamefont {Liberto}, \citenamefont {Goldman}, \citenamefont
  {Bloch},\ and\ \citenamefont {Aidelsburger}}]{Wintersperger20_NatPhys}%
  \BibitemOpen
  \bibfield  {author} {\bibinfo {author} {\bibfnamefont {Karen}\ \bibnamefont
  {Wintersperger}}, \bibinfo {author} {\bibfnamefont {Christoph}\ \bibnamefont
  {Braun}}, \bibinfo {author} {\bibfnamefont {F.~Nur}\ \bibnamefont {\"Unal}},
  \bibinfo {author} {\bibfnamefont {Andre}\ \bibnamefont {Eckardt}}, \bibinfo
  {author} {\bibfnamefont {Marco~Di}\ \bibnamefont {Liberto}}, \bibinfo
  {author} {\bibfnamefont {Nathan}\ \bibnamefont {Goldman}}, \bibinfo {author}
  {\bibfnamefont {Immanuel}\ \bibnamefont {Bloch}}, \ and\ \bibinfo {author}
  {\bibfnamefont {Monika}\ \bibnamefont {Aidelsburger}},\ }\bibfield  {title}
  {\enquote {\bibinfo {title} {Realization of an anomalous {F}loquet
  topological system with ultracold atoms},}\ }\href {\doibase
  10.1038/s41567-020-0949-y} {\bibfield  {journal} {\bibinfo  {journal} {Nat.
  Phys.}\ }\textbf {\bibinfo {volume} {16}},\ \bibinfo {pages} {1058} (\bibinfo
  {year} {2020})}\BibitemShut {NoStop}%
\bibitem [{\citenamefont {\"Unal}\ \emph
  {et~al.}(2019{\natexlab{b}})\citenamefont {\"Unal}, \citenamefont
  {Seradjeh},\ and\ \citenamefont {Eckardt}}]{Unal19_PRL}%
  \BibitemOpen
  \bibfield  {author} {\bibinfo {author} {\bibfnamefont {F.~Nur}\ \bibnamefont
  {\"Unal}}, \bibinfo {author} {\bibfnamefont {Babak}\ \bibnamefont
  {Seradjeh}}, \ and\ \bibinfo {author} {\bibfnamefont {Andr\'e}\ \bibnamefont
  {Eckardt}},\ }\bibfield  {title} {\enquote {\bibinfo {title} {How to directly
  measure floquet topological invariants in optical lattices},}\ }\href
  {\doibase 10.1103/PhysRevLett.122.253601} {\bibfield  {journal} {\bibinfo
  {journal} {Phys. Rev. Lett.}\ }\textbf {\bibinfo {volume} {122}},\ \bibinfo
  {pages} {253601} (\bibinfo {year} {2019}{\natexlab{b}})}\BibitemShut
  {NoStop}%
\bibitem [{\citenamefont {Tarnowski}\ \emph {et~al.}(2019)\citenamefont
  {Tarnowski}, \citenamefont {{\"U}nal}, \citenamefont {Flaschner},
  \citenamefont {Rem}, \citenamefont {Eckardt}, \citenamefont {Sengstock},\
  and\ \citenamefont {Weitenberg}}]{Tarnowski19_NatCom}%
  \BibitemOpen
  \bibfield  {author} {\bibinfo {author} {\bibfnamefont {Matthias}\
  \bibnamefont {Tarnowski}}, \bibinfo {author} {\bibfnamefont {F.~Nur}\
  \bibnamefont {{\"U}nal}}, \bibinfo {author} {\bibfnamefont {Nick}\
  \bibnamefont {Flaschner}}, \bibinfo {author} {\bibfnamefont {Benno~S}\
  \bibnamefont {Rem}}, \bibinfo {author} {\bibfnamefont {Andre}\
  \bibnamefont {Eckardt}}, \bibinfo {author} {\bibfnamefont {Klaus}\
  \bibnamefont {Sengstock}}, \ and\ \bibinfo {author} {\bibfnamefont
  {Christof}\ \bibnamefont {Weitenberg}},\ }\bibfield  {title} {\enquote
  {\bibinfo {title} {Measuring topology from dynamics by obtaining the {C}hern
  number from a linking number},}\ }\href
  {https://www.nature.com/articles/s41467-019-09668-y} {\bibfield  {journal}
  {\bibinfo  {journal} {Nat. Commun.}\ }\textbf {\bibinfo {volume} {10}},\
  \bibinfo {pages} {1728} (\bibinfo {year} {2019})}\BibitemShut {NoStop}%
\bibitem [{\citenamefont {Maczewsky}\ \emph {et~al.}(2017)\citenamefont
  {Maczewsky}, \citenamefont {Zeuner}, \citenamefont {Nolte},\ and\
  \citenamefont {Szameit}}]{Maczewsky17_NatCommun}%
  \BibitemOpen
  \bibfield  {author} {\bibinfo {author} {\bibfnamefont {Lukas~J.}\
  \bibnamefont {Maczewsky}}, \bibinfo {author} {\bibfnamefont {Julia~M.}\
  \bibnamefont {Zeuner}}, \bibinfo {author} {\bibfnamefont {Stefan}\
  \bibnamefont {Nolte}}, \ and\ \bibinfo {author} {\bibfnamefont {Alexander}\
  \bibnamefont {Szameit}},\ }\bibfield  {title} {\enquote {\bibinfo {title}
  {Observation of photonic anomalous {F}loquet topological insulators},}\
  }\href {\doibase 10.1038/ncomms13756} {\bibfield  {journal} {\bibinfo
  {journal} {Nat. Commun.}\ }\textbf {\bibinfo {volume} {8}} (\bibinfo {year}
  {2017}),\ 10.1038/ncomms13756}\BibitemShut {NoStop}%
\bibitem [{\citenamefont {Mukherjee}\ \emph {et~al.}(2017)\citenamefont
  {Mukherjee}, \citenamefont {Spracklen}, \citenamefont {Valiente},
  \citenamefont {Andersson}, \citenamefont {Ohberg}, \citenamefont {Goldman},\
  and\ \citenamefont {Thomson}}]{Mukherjee17_NatComm}%
  \BibitemOpen
  \bibfield  {author} {\bibinfo {author} {\bibfnamefont {Sebabrata}\
  \bibnamefont {Mukherjee}}, \bibinfo {author} {\bibfnamefont {Alexander}\
  \bibnamefont {Spracklen}}, \bibinfo {author} {\bibfnamefont {Manuel}\
  \bibnamefont {Valiente}}, \bibinfo {author} {\bibfnamefont {Erika}\
  \bibnamefont {Andersson}}, \bibinfo {author} {\bibfnamefont {Patrik}\
  \bibnamefont {Ohberg}}, \bibinfo {author} {\bibfnamefont {Nathan}\
  \bibnamefont {Goldman}}, \ and\ \bibinfo {author} {\bibfnamefont {Robert~R}\
  \bibnamefont {Thomson}},\ }\bibfield  {title} {\enquote {\bibinfo {title}
  {Experimental observation of anomalous topological edge modes in a slowly
  driven photonic lattice},}\ }\href {\doibase 10.1038/ncomms13918} {\bibfield
  {journal} {\bibinfo  {journal} {Nat. Commun.}\ }\textbf {\bibinfo {volume}
  {8}} (\bibinfo {year} {2017}),\ 10.1038/ncomms13918}\BibitemShut {NoStop}%
\bibitem [{\citenamefont {Du}\ \emph {et~al.}(2017)\citenamefont {Du},
  \citenamefont {Zhou},\ and\ \citenamefont {Fiete}}]{DuFiete17_PRB}%
  \BibitemOpen
  \bibfield  {author} {\bibinfo {author} {\bibfnamefont {Liang}\ \bibnamefont
  {Du}}, \bibinfo {author} {\bibfnamefont {Xiaoting}\ \bibnamefont {Zhou}}, \
  and\ \bibinfo {author} {\bibfnamefont {Gregory~A.}\ \bibnamefont {Fiete}},\
  }\bibfield  {title} {\enquote {\bibinfo {title} {Quadratic band touching
  points and flat bands in two-dimensional topological floquet systems},}\
  }\href {\doibase 10.1103/PhysRevB.95.035136} {\bibfield  {journal} {\bibinfo
  {journal} {Phys. Rev. B}\ }\textbf {\bibinfo {volume} {95}},\ \bibinfo
  {pages} {035136} (\bibinfo {year} {2017})}\BibitemShut {NoStop}%
\bibitem [{\citenamefont {Lignier}\ \emph {et~al.}(2007)\citenamefont
  {Lignier}, \citenamefont {Sias}, \citenamefont {Ciampini}, \citenamefont
  {Singh}, \citenamefont {Zenesini}, \citenamefont {Morsch},\ and\
  \citenamefont {Arimondo}}]{Lignier07_PRL_dynFrustr}%
  \BibitemOpen
  \bibfield  {author} {\bibinfo {author} {\bibfnamefont {H.}~\bibnamefont
  {Lignier}}, \bibinfo {author} {\bibfnamefont {C.}~\bibnamefont {Sias}},
  \bibinfo {author} {\bibfnamefont {D.}~\bibnamefont {Ciampini}}, \bibinfo
  {author} {\bibfnamefont {Y.}~\bibnamefont {Singh}}, \bibinfo {author}
  {\bibfnamefont {A.}~\bibnamefont {Zenesini}}, \bibinfo {author}
  {\bibfnamefont {O.}~\bibnamefont {Morsch}}, \ and\ \bibinfo {author}
  {\bibfnamefont {E.}~\bibnamefont {Arimondo}},\ }\bibfield  {title} {\enquote
  {\bibinfo {title} {Dynamical control of matter-wave tunneling in periodic
  potentials},}\ }\href {\doibase 10.1103/PhysRevLett.99.220403} {\bibfield
  {journal} {\bibinfo  {journal} {Phys. Rev. Lett.}\ }\textbf {\bibinfo
  {volume} {99}},\ \bibinfo {pages} {220403} (\bibinfo {year}
  {2007})}\BibitemShut {NoStop}%
\bibitem [{\citenamefont {Zak}(1989)}]{Zak2}%
  \BibitemOpen
  \bibfield  {author} {\bibinfo {author} {\bibfnamefont {J.}~\bibnamefont
  {Zak}},\ }\bibfield  {title} {\enquote {\bibinfo {title} {Berry's phase for
  energy bands in solids},}\ }\href {\doibase 10.1103/PhysRevLett.62.2747}
  {\bibfield  {journal} {\bibinfo  {journal} {Phys. Rev. Lett.}\ }\textbf
  {\bibinfo {volume} {62}},\ \bibinfo {pages} {2747--2750} (\bibinfo {year}
  {1989})}\BibitemShut {NoStop}%
\bibitem [{\citenamefont {Brown}\ \emph {et~al.}(2021)\citenamefont {Brown},
  \citenamefont {Chang}, \citenamefont {Schwarz}, \citenamefont {Leung},
  \citenamefont {Kozii}, \citenamefont {Avdoshkin}, \citenamefont {Moore},\
  and\ \citenamefont {Stamper-Kurn}}]{Kurn21_arXiv_bandsingularity}%
  \BibitemOpen
  \bibfield  {author} {\bibinfo {author} {\bibfnamefont {Charles~D.}\
  \bibnamefont {Brown}}, \bibinfo {author} {\bibfnamefont {Shao-Wen}\
  \bibnamefont {Chang}}, \bibinfo {author} {\bibfnamefont {Malte~N.}\
  \bibnamefont {Schwarz}}, \bibinfo {author} {\bibfnamefont {Tsz-Him}\
  \bibnamefont {Leung}}, \bibinfo {author} {\bibfnamefont {Vladyslav}\
  \bibnamefont {Kozii}}, \bibinfo {author} {\bibfnamefont {Alexander}\
  \bibnamefont {Avdoshkin}}, \bibinfo {author} {\bibfnamefont {Joel~E.}\
  \bibnamefont {Moore}}, \ and\ \bibinfo {author} {\bibfnamefont {Dan}\
  \bibnamefont {Stamper-Kurn}},\ }\bibfield  {title} {\enquote {\bibinfo
  {title} {Direct geometric probe of singularities in band structure},}\ }\href
  {\doibase https://doi.org/10.48550/arXiv.2109.03354} {\bibfield  {journal}
  {\bibinfo  {journal} {ArXiv e-prints}\ } (\bibinfo {year} {2021}),\
  https://doi.org/10.48550/arXiv.2109.03354},\ \Eprint
  {http://arxiv.org/abs/2109.03354} {arXiv:2109.03354} \BibitemShut {NoStop}%
\bibitem [{\citenamefont {Leung}\ \emph {et~al.}(2020)\citenamefont {Leung},
  \citenamefont {Schwarz}, \citenamefont {Chang}, \citenamefont {Brown},
  \citenamefont {Unnikrishnan},\ and\ \citenamefont
  {Stamper-Kurn}}]{LeungStamperKurn20_PRL_kgm}%
  \BibitemOpen
  \bibfield  {author} {\bibinfo {author} {\bibfnamefont {Tsz-Him}\ \bibnamefont
  {Leung}}, \bibinfo {author} {\bibfnamefont {Malte~N.}\ \bibnamefont
  {Schwarz}}, \bibinfo {author} {\bibfnamefont {Shao-Wen}\ \bibnamefont
  {Chang}}, \bibinfo {author} {\bibfnamefont {Charles~D.}\ \bibnamefont
  {Brown}}, \bibinfo {author} {\bibfnamefont {Govind}\ \bibnamefont
  {Unnikrishnan}}, \ and\ \bibinfo {author} {\bibfnamefont {Dan}\ \bibnamefont
  {Stamper-Kurn}},\ }\bibfield  {title} {\enquote {\bibinfo {title}
  {Interaction-enhanced group velocity of bosons in the flat band of an optical
  kagome lattice},}\ }\href {\doibase 10.1103/PhysRevLett.125.133001}
  {\bibfield  {journal} {\bibinfo  {journal} {Phys. Rev. Lett.}\ }\textbf
  {\bibinfo {volume} {125}},\ \bibinfo {pages} {133001} (\bibinfo {year}
  {2020})}\BibitemShut {NoStop}%
\bibitem [{\citenamefont {Wang}\ \emph {et~al.}(2013)\citenamefont {Wang},
  \citenamefont {Steinberg}, \citenamefont {Jarillo-Herrero},\ and\
  \citenamefont {Gedik}}]{wang2013observation}%
  \BibitemOpen
  \bibfield  {author} {\bibinfo {author} {\bibfnamefont {YH}~\bibnamefont
  {Wang}}, \bibinfo {author} {\bibfnamefont {Hadar}\ \bibnamefont {Steinberg}},
  \bibinfo {author} {\bibfnamefont {Pablo}\ \bibnamefont {Jarillo-Herrero}}, \
  and\ \bibinfo {author} {\bibfnamefont {Nuh}\ \bibnamefont {Gedik}},\
  }\bibfield  {title} {\enquote {\bibinfo {title} {Observation of floquet-bloch
  states on the surface of a topological insulator},}\ }\href@noop {}
  {\bibfield  {journal} {\bibinfo  {journal} {Science}\ }\textbf {\bibinfo
  {volume} {342}},\ \bibinfo {pages} {453--457} (\bibinfo {year}
  {2013})}\BibitemShut {NoStop}%
\bibitem [{\citenamefont {Trevisan}\ \emph {et~al.}(2022)\citenamefont
  {Trevisan}, \citenamefont {Arribi}, \citenamefont {Heinonen}, \citenamefont
  {Slager},\ and\ \citenamefont {Orth}}]{Threvisan2022}%
  \BibitemOpen
  \bibfield  {author} {\bibinfo {author} {\bibfnamefont {Tha\'{\i}s~V.}\
  \bibnamefont {Trevisan}}, \bibinfo {author} {\bibfnamefont {Pablo~Villar}\
  \bibnamefont {Arribi}}, \bibinfo {author} {\bibfnamefont {Olle}\ \bibnamefont
  {Heinonen}}, \bibinfo {author} {\bibfnamefont {Robert-Jan}\ \bibnamefont
  {Slager}}, \ and\ \bibinfo {author} {\bibfnamefont {Peter~P.}\ \bibnamefont
  {Orth}},\ }\bibfield  {title} {\enquote {\bibinfo {title} {Bicircular light
  floquet engineering of magnetic symmetry and topology and its application to
  the dirac semimetal ${\mathrm{cd}}_{3}{\mathrm{as}}_{2}$},}\ }\href {\doibase
  10.1103/PhysRevLett.128.066602} {\bibfield  {journal} {\bibinfo  {journal}
  {Phys. Rev. Lett.}\ }\textbf {\bibinfo {volume} {128}},\ \bibinfo {pages}
  {066602} (\bibinfo {year} {2022})}\BibitemShut {NoStop}%
\bibitem [{\citenamefont {Milnor}\ and\ \citenamefont
  {Stasheff}(1974)}]{Milnor:1974}%
  \BibitemOpen
  \bibfield  {author} {\bibinfo {author} {\bibfnamefont {John~W.}\ \bibnamefont
  {Milnor}}\ and\ \bibinfo {author} {\bibfnamefont {James~D.}\ \bibnamefont
  {Stasheff}},\ }\href@noop {} {\emph {\bibinfo {title} {{C}haracteristic
  classes}}}\ (\bibinfo  {publisher} {Princeton University Press},\ \bibinfo
  {address} {Princeton, New Jersey},\ \bibinfo {year} {1974})\BibitemShut
  {NoStop}%
\bibitem [{\citenamefont {Hatcher}(2003)}]{Hatcher_2}%
  \BibitemOpen
  \bibfield  {author} {\bibinfo {author} {\bibfnamefont {A.}~\bibnamefont
  {Hatcher}},\ }\href@noop {} {\emph {\bibinfo {title} {{V}ector {B}undles and
  {K}-{T}heory}}}\ (\bibinfo  {publisher} {Unpublished},\ \bibinfo {year}
  {2003})\BibitemShut {NoStop}%
\bibitem [{\citenamefont {Zak}(1979)}]{Zak1}%
  \BibitemOpen
  \bibfield  {author} {\bibinfo {author} {\bibfnamefont {J.}~\bibnamefont
  {Zak}},\ }\bibfield  {title} {\enquote {\bibinfo {title} {Lattice
  representations in solids},}\ }\href {\doibase 10.1103/PhysRevB.20.2228}
  {\bibfield  {journal} {\bibinfo  {journal} {Phys. Rev. B}\ }\textbf {\bibinfo
  {volume} {20}},\ \bibinfo {pages} {2228--2237} (\bibinfo {year}
  {1979})}\BibitemShut {NoStop}%
\bibitem [{\citenamefont {Su}\ \emph {et~al.}(1979)\citenamefont {Su},
  \citenamefont {Schrieffer},\ and\ \citenamefont {Heeger}}]{solitonsZak}%
  \BibitemOpen
  \bibfield  {author} {\bibinfo {author} {\bibfnamefont {W.~P.}\ \bibnamefont
  {Su}}, \bibinfo {author} {\bibfnamefont {J.~R.}\ \bibnamefont {Schrieffer}},
  \ and\ \bibinfo {author} {\bibfnamefont {A.~J.}\ \bibnamefont {Heeger}},\
  }\bibfield  {title} {\enquote {\bibinfo {title} {Solitons in
  polyacetylene},}\ }\href {\doibase 10.1103/PhysRevLett.42.1698} {\bibfield
  {journal} {\bibinfo  {journal} {Phys. Rev. Lett.}\ }\textbf {\bibinfo
  {volume} {42}},\ \bibinfo {pages} {1698--1701} (\bibinfo {year}
  {1979})}\BibitemShut {NoStop}%
\end{thebibliography}%

\end{document}